\documentclass[11pt]{article}
\baselineskip
\usepackage{amssymb}
\usepackage{graphicx}
\usepackage{latexsym,cite}
\usepackage{epsfig}
\usepackage{amsmath, bbm, color}
\usepackage{hyperref}

\newcommand{\SU}{\mathrm{SU}}

\newcommand{\R}{\mathbb{R}}

\newcommand{\re}{{\rm{Re}}}
\newcommand{\dd}{{\rm{d}}}
\newcommand{\Tr}{{\rm Tr\,}}
\newcommand{\MP}{M_{\tiny\mbox{P}}}
\newcommand{\VS}{\mathcal{V}_{\tiny\mbox{S}}}
\newcommand{\nconf}{n_{\mbox{\tiny{conf}}}}
\newcommand{\Se}{S^{\tiny\mbox{E}}}

\newcommand{\Sel}{S^{\tiny\mbox{E}}_{\tiny\mbox{L}}}
\newcommand{\gsqeff}{g^2_{\tiny\mbox{eff}}}
\newcommand{\Zlat}{Z_{\tiny\mbox{L}}}
\newcommand{\eq}{\begin{equation}}
\newcommand{\en}{\end{equation}}
\newcommand{\ba}{\begin{eqnarray}}
\newcommand{\ea}{\end{eqnarray}}
\newcommand{\arsinh}{\mathrm{arsinh}}
\newcommand{\be}{\begin{equation}}
\newcommand{\ee}{\end{equation}}
\newcommand{\eea}{\end{eqnarray}}
\newcommand{\bea}{\begin{eqnarray}}
\newcommand{\n}{\nu}
\newcommand{\cA}{{\cal A}}
\newcommand{\cf}{{\bf f}}
\newcommand{\ct}{\tilde{\Delta}}
\newcommand{\tT}{\tilde{T}}
\newcommand{\cO}{{\cal O}}
\newcommand{\f}{\phi}
\newcommand{\m}{\mu}
\newcommand{\La}{\Lambda}
\newcommand{\lab}{\label}
\newcommand{\lef}{\left}
\newcommand{\ri}{\right}
\newcommand{\6}{\partial}
\newcommand{\bz}{\begin{itemize}}
\newcommand{\ez}{\end{itemize}}

\begin{document}

\begin{titlepage}
\vskip0.5cm
\begin{flushright}
DFTT 30/11\\
CERN-PH-TH-2011-266\\
HIP-2011-28/TH\\
\end{flushright}
\vskip0.5cm
\begin{center}
{\Large\bf
Thermodynamics of $\SU(N)$ Yang-Mills theories in $2+1$ dimensions II -- The deconfined phase
}
\end{center}
\vskip1.3cm
\begin{center}
Michele~Caselle$^{a}$, Luca~Castagnini$^{b}$, Alessandra~Feo$^{c}$, Ferdinando~Gliozzi$^{a}$,\\ Umut~G\"ursoy$^{d}$, Marco~Panero$^{e}$ and Andreas~Sch\"afer$^{b}$
\end{center}
\vskip1.5cm
\centerline{\sl  $^a$ Dipartimento di Fisica Teorica dell'Universit\`a di Torino and INFN, Sezione di Torino,}
 \centerline{\sl Via P.~Giuria 1, I-10125 Torino, Italy}
\vskip0.5cm
\centerline{\sl  $^b$ Institute for Theoretical Physics,  University of Regensburg,}
 \centerline{\sl D-93040 Regensburg, Germany}
\vskip0.5cm
\centerline{\sl  $^c$ Dipartimento di Fisica, Universit\`a di Parma,}
 \centerline{\sl  Viale G.P. Usberti 7/A, I-43124 Parma, Italy}
\vskip0.5cm
\centerline{\sl  $^d$ Theory Group, Physics Department, CERN,}
 \centerline{\sl CH-1211 Geneva 23, Switzerland}
\vskip0.5cm
\centerline{\sl  $^e$ Department of Physics and Helsinki Institute of Physics, University of Helsinki,}
 \centerline{\sl FIN-00014 Helsinki, Finland}
\vskip0.5cm
\begin{center}
{\sl  E-mail:} \hskip 5mm \texttt{caselle@to.infn.it, luca.castagnini@physik.uni-regensburg.de, alessandra.feo@fis.unipr.it, gliozzi@to.infn.it, umut.gursoy@cern.ch, marco.panero@helsinki.fi, andreas.schaefer@physik.uni-regensburg.de}
\end{center}
\vskip1.0cm
\begin{abstract}
We present a non-perturbative study of the equation of state in the deconfined phase of Yang-Mills theories in $D=2+1$ dimensions. We introduce a holographic model, based on the improved holographic QCD model, from which we derive a non-trivial relation between the order of the deconfinement phase transition and the behavior of the trace of the energy-momentum tensor as a function of the temperature $T$. We compare the theoretical predictions of this holographic model with a new set of high-precision numerical results from lattice simulations of $\SU(N)$ theories with $N=2$, $3$, $4$, $5$ and $6$ colors. The latter reveal that, similarly to the $D=3+1$ case, the bulk equilibrium thermodynamic quantities (pressure, trace of the energy-momentum tensor, energy density and entropy density) exhibit nearly perfect proportionality to the number of gluons, and can be successfully compared with the holographic predictions in a broad range of temperatures. Finally, we also show that, again similarly to the $D=3+1$ case, the trace of the energy-momentum tensor appears to be proportional to $T^2$ in a wide temperature range, starting from approximately $1.2~T_c$, where $T_c$ denotes the critical deconfinement temperature.
\end{abstract}
\vspace*{0.2cm}

\end{titlepage}

\section{Introduction and motivation}
\label{introsect}

The phase diagram of strongly interacting matter has been studied in an extensive experimental program since the 1980's. During the last decade, the main heavy ion collision facilities have provided convincing evidence for the existence of a new state of matter, which is qualitatively different from usual hadronic matter, and appears to behave as a nearly ideal fluid~\cite{finiteTexperiments}. While these results confirm the intuitive theoretical expectation that, at high temperatures or densities, asymptotic freedom leads to deconfinement, i.e. to the liberation of colored particles from hadrons~\cite{first_deconfinement_prediction}, they also reveal that the deconfined plasma is quite different from a gas of nearly free quarks and gluons, and should be rather described as a strongly coupled fluid. This makes the theoretical description of the equation of state of this system at temperatures close to deconfinement particularly challenging: weak-coupling computations~\cite{perturbative_computations} have to be pushed to high orders, but the convergence of perturbative expansions in thermal gauge theories is generally poor, and the evaluation of high-order terms is complicated by the appearence of severe infrared (IR) divergences~\cite{Linde_problem}. The latter reveal the mathematically non-trivial structure of perturbative expansions in finite-temperature QCD (with terms which are non-analytical in $\alpha_s$), and are related to the existence of an ultra-soft, chromomagnetic energy scale, which retains an intrinsically non-perturbative nature, and to long-wavelength modes that are strongly coupled at \emph{all} temperatures.

As a consequence, numerical computations on the lattice are the main tool to derive the predictions of QCD at the temperatures probed in experiments, and in the last few years, various collaborations have presented results for the QCD equation of state at vanishing chemical potential, obtained from simulations including dynamical quarks at or close to the physical point~\cite{finite_T_lattice_reviews}.\footnote{By contrast, the progress in lattice simulations of the QCD equation of state at finite net baryon density has been slower, due to the existence of a severe sign problem~\cite{finite_density_lattice_QCD}. As a consequence, in this case one often obtains useful insight from the numerical study of appropriate effective models, see, e.g., ref.~\cite{DePietri:2007ak}.} At the same time, high-precision lattice results have also been obtained for various equilibrium thermodynamic properties in $\SU(N)$ Yang-Mills theories with a large number of colors $N$, which is a particularly interesting limit, for several reasons.\footnote{Some of the surprising mathematical properties arising in the large-$N$ limit of a generic quantum theory are related to the fact that, generally, the large-$N$ limit can be interpreted as a sort of ``classical limit''. The meaning of this statement is made precise in ref.~\cite{Yaffe:1981vf}, with the definition of an appropriate basis of coherent states and a classical Hamiltonian. The construction, however, can be carried out explicitly, and leads to an exact solution, only for certain particularly simple models.} First of all, the large-$N$ limit at fixed 't~Hooft coupling $\lambda=g^2 N$ and fixed number of flavors $N_f$~\cite{'tHooftlargeN} provides a natural interpretation for some non-trivial features of QCD (such as, for instance, the OZI rule~\cite{OZI_rule}), and leads to a topological classification of Feynman diagrams, in which the dominant contributions come from planar graphs. This is suggestive of an analogy with similar expansions in closed string theory~\cite{Aharony:1999ti}. Moreover, for $N \to \infty$ one expects that all correlation functions of gauge-invariant operators factorize, and that the functional integral describing a large-$N$ gauge theory should be dominated by a single ``master'' gauge field~\cite{Gopakumar:1994iq}; the translational invariance properties of the latter are related to the ideas of large-$N$ volume independence~\cite{volume_reduction}. 

As it concerns the phase diagram of QCD-like theories, it is interesting to note that the large-$N$ limit leads to interesting implications at finite density, including, in particular, a possible ``quarkyonic phase''~\cite{McLerran:2007qj}. Moreover, this limit is also important for applications of the conjectured correspondence between gauge and string theories~\cite{Maldacena_conjecture} to study the strongly interacting plasma~\cite{gauge_string_applications_to_QCD_plasma}, since these computations are done in the infinite-$N$ limit.

For these reasons, it is interesting to understand how much the thermal properties of strongly interacting gauge theories depend on the number of colors. Recent lattice simulations of large-$N$ Yang-Mills theories at finite temperature~\cite{finiteTlargeNlatticeresults, Panero:2009tv} have revealed that, in the deconfined phase, the bulk thermodynamic observables are essentially independent of $N$ (except for a trivial proportionality to the number of gluons), and that $\SU(3)$~\cite{SU3_EoS} is close to the large-$N$ limit.\footnote{Similar findings have been obtained from large-$N$ simulations at zero temperature, see, e.g., ref.~\cite{Teper:2009uf} and references therein.} This result is particularly interesting from a theoretical point of view, since it provides  support to analytical studies of the QCD plasma based on the approximation of an infinite number of colors, and, furthermore, it can shed light onto the effective degrees of freedom relevant for the plasma near deconfinement, and/or rule out possible effective models.

A different perspective on the hot QCD plasma is based on the study of non-Abelian gauge theories in a lower-dimensional spacetime. While the case of $D=1+1$ dimensions is essentially trivial~\cite{Gross:1980he}, in $D=2+1$ these theories are characterized by rich dynamics~\cite{Teper:1998te}, and share many qualitative features with their $D=3+1$ analogues. In particular, at low energies they are characterized by a spectrum of color-singlet states with a finite mass-gap and by linear confinement, and they exhibit a deconfining transition at a finite temperature $T_c$. Moreover, these theories can also be studied using techniques inspired by the AdS/CFT correspondence: comparing the results obtained from first-principle lattice computations with those derived from the gauge/string duality can provide a useful test-bed for the application of holographic methods to study strongly coupled systems in $D=2+1$ dimensions, such as those relevant for condensed matter systems at criticality~\cite{Sachdev:2010ch}. Finally, one further motivation to look at the Yang-Mills equation of state in $2+1$ dimensions stems from the observation that, in $D=3+1$, the trace of the energy-momentum tensor in the deconfined phase appears to be proportional to $T^2$ over a broad range of temperatures~\cite{Panero:2009tv,SU3_EoS,Tsquare_in_4D}. This behavior seems to be at odds with the expectation from perturbative computations, which would rather predict a logarithmic dependence on the temperature. In order to understand the physical origin of this characteristic behavior in the physical case of $D=3+1$, it is instructive to investigate whether the same phenomenon also occurs for a generic $\SU(N)$ Yang-Mills theory in the lower-dimensional setup, given that the theories in $3+1$ and in $2+1$ dimensions have both some similarities and some obvious qualitative differences.

For these reasons, in this work we present a systematic study of finite-temperature $\SU(N)$ Yang-Mills theories in $D=2+1$ dimensions. Having already discussed the confining phase of these theories in a previous work~\cite{Caselle:2011fy}, in the present article we focus on the equilibrium thermodynamic properties in the deconfined phase. In particular, we investigate the strongly coupled regime (close to the deconfinement temperature $T_c$), where physical quantities cannot be reliably caculated via weak-coupling expansions, and compare the theoretical results obtained from two different non-perturbative approaches: holographic computations based on the gauge/string correspondence, and numerical simulations in the lattice regularization. 

The structure of this paper is as follows. First, in section~\ref{sec:holographic_model} we construct a holographic model, which is expected to describe the deconfined finite-temperature phase of strongly coupled $\SU(N)$ gauge theories in $D=2+1$. Then, in section~\ref{sec:2+1_gauge_theories}, we review the basic properties of $\SU(N)$ Yang-Mills theories in $2+1$ spacetime dimensions, and introduce their regularization on a Euclidean lattice. Next, in section~\ref{sec:results} we present a set of high-precision results from our numerical lattice simulations of these theories, for different values of $N$
ranging from $2$ to $6$. After discussing the extrapolation to the thermodynamic and continuum limits, we investigate the dependence of the pressure ($p$), of the trace of the energy-momentum tensor (or interaction measure, denoted by $\Delta$), and of the energy ($\epsilon$) and entropy ($s$) densities on the number of colors, and compare the prediction for $\Delta$ to the holographic model. Section~\ref{sec:conclusions} includes a discussion of our findings and their implications. The appendix~\ref{app:lattice_SB} reports the details of the computation of the lattice Stefan-Boltzmann limit in $D=2+1$ and $D=3+1$ dimensions.

\section{A holographic model}
\label{sec:holographic_model}

\subsection{Generalities}
The gauge-gravity correspondence (``holography" for short)~\cite{Maldacena_conjecture} successfully reproduces most of the salient features of large-$N$ gauge theories. Although the first examples of holography involved supersymmetric and conformal quantum field theories, the correspondence was soon generalized to more realistic examples, including theories with linear confinement and no supersymmetry~\cite{Witten2}, hence in the same class as QCD. However, these models (sometimes referred to as models built in a ``top-down approach"), that stem from D-brane constructions in type IIA or IIB string theory, generally have an infinite number of undesired scalar operators in their spectrum, arising from the Kaluza-Klein modes on the internal extra-dimensions of the ten-dimensional parent theory.

In the meantime, an alternative ``bottom-up" holographic approach has been developed~\cite{pheno}: it uses minimal ingredients to model the desired features of confining gauge theories on the gravity side, in a more direct and ``economic'' fashion. Our approach here consists of an advanced version of the bottom-up construction, that is known as ``improved holographic QCD" (IHQCD)~\cite{GK,GKN}---see ref.~\cite{Kiritsis} for a review, and ref.~\cite{Gubser} for similar constructions in the literature. In what follows, we first explain and review the setup of IHQCD.

Holography generally associates the energy dependence of the field theory with a radial direction $r$ perpendicular to the $D$ dimensions of the Minkowski spacetime in which the gauge theory is defined. Therefore the most economic ``bottom-up'' approach to a $D$-dimensional QFT involves a $(D+1)$-dimensional gravitational background. In addition, the holographic correspondence associates a bulk field to each operator that is relevant or marginal in the IR.\footnote{Typically, the holographic description in bottom-up constructions is only reliable in the IR of the field theory, because the far UV region of the background suffers from large curvature corrections. However, the reliable region turns out to be quite large in IHQCD models~\cite{GK}.} For pure Yang-Mills theory, there are two such marginal operators: the energy-momentum tensor $T_{\m\n}$ and the gluon operator $\Tr F^2/\La^{4-D}$ where $\La$ is the dynamically generated energy scale of the theory (for $D \le 4$). The bulk fields that are dual to these operators are the metric $g_{\m\n}(r)$ and the dilaton field $\f(r)$. The dependence of these fields on the radial coordinate $r$ corresponds to the renormalization group scale dependence of the corresponding operators in the field theory. In particular, the profile of $\f(r)$ encodes how the (dimensionless) coupling constant $\gsqeff = g^2/\La^{4-D}$ runs with the energy.\footnote{This is so, because in string theory the field $e^{\f}$ couples to the operator $\Tr F^2/\La^{4-D}$ in the same way as the dimensionless coupling $g^2/\La^{4-D}$ does.} On the other hand, in order to make $\f(r)$ run with $r$ in a dynamical gravitational setup, one needs to turn on a potential for it. Therefore, the minimal general relativity action of the IHQCD is:
%%%%%%%%%%%%%%%%%%%%%%%%%%%%%%%%%%%%%%%
\begin{equation}
\label{action}
 {\cal A} = \MP^{D-1}N^2 \int d^{D+1}x \sqrt{-g}\left[ R - \xi (\6\f)^2 - \mathcal{V}(\f)
  \right]+\cdots
\end{equation}
%%%%%%%%%%%%%%%%%%%%%%%%%%%%%%%%%%%%%%%
(where the ellipsis denotes some boundary counter-terms that should be introduced to render the variational problem on geometries with a boundary well-defined; we will not need the explicit form of these terms here). The coefficient $\xi$ is an unspecified normalization constant\footnote{This coefficient was set to $4/3$ in ref.~\cite{GK}, as motivated by embedding the theory in non-critical string theory.} that will not play an important r\^ole in what follows. In particular, it can be absorbed into $\f$ by a redefinition. Note that the action is proportional to $N^2$ and to a positive power of $\MP$, which denotes a ``reduced'' Planck mass,\footnote{The actual Planck mass is the entire expression $\MP N^{2/(D-1)}$.} thus, in the large-$N$ limit of the gauge theory, gravitational interactions are suppressed. For convenience, we keep the normalization of the scalar kinetic term unspecified, except that we assume $\xi >0$.  We also assume that the scalar potential has a single AdS minimum, that corresponds to the UV limit of the dual field theory:
%%%%%%%%%%%%%%%%%%%%%%%%%%%%%%%%%%%%%%%
\be\lab{min}
\mathcal{V}^{\,\prime}(\f)\bigg|_{\f=\f_{UV}} = 0, \qquad \mathcal{V}(\f_{UV}) = \frac{D(D-1)}{\ell^2},
\ee
%%%%%%%%%%%%%%%%%%%%%%%%%%%%%%%%%%%%%%%
where $\ell$ is the AdS length scale.

\paragraph{Vacuum solution:} The solution to the action in eq.~(\ref{action}) that corresponds to the {\em vacuum} of the dual field theory is of the form:
%%%%%%%%%%%%%%%%%%%%%%%%%%%%%%%%%%%%%%%
\begin{equation}\label{Vac}
  ds^2 = b^2_0(r)\lef( dr^2 + dx_{D-1}^2 - dt^2 \ri), \qquad \f=
  \f_0(r).
\end{equation}
%%%%%%%%%%%%%%%%%%%%%%%%%%%%%%%%%%%%%%%
Here the scale factor of the metric $b_0(r)$ is of the AdS form $b_0(r) = \ell/r$ only if the dilaton potential is constant: $\mathcal{V}= D(D-1)/\ell^2$. More generally, when $\mathcal{V}$ is a non-trivial function of $\f$, the function $b_0$ attains the AdS form only in the UV, i.e. for $ r\to 0$, while it deviates from AdS in the IR limit, i.e. for $r\to \infty$. Its profile can be determined by solving the Einstein's equations, given the potential $\mathcal{V}(\f)$. The IR part of the geometry characterizes the confinement properties in the vacuum of the dual field theory. In ref.~\cite{GKN}, the various confining asymptotics were classified. In particular, we shall be interested in the {\em confining geometries} (with gapped and discrete spectrum) of the form:
%%%%%%%%%%%%%%%%%%%%%%%%%%%%%%%%%%%%%%%
\begin{equation}\label{IR}
b_0(r)\to e^{-(r\La)^\alpha + \cdots}, \qquad \alpha>1 \qquad \mbox{for}\,\,\, r\to\infty,
\end{equation}
%%%%%%%%%%%%%%%%%%%%%%%%%%%%%%%%%%%%%%%
where the integration constant $\La$ corresponds to the dynamically generated energy scale, and the ellipsis denotes sub-leading terms that are typically logarithmic in $r$~\cite{GKN}. The parameter $\alpha$ is directly related to the large-$\f$ asymptotics of the dilaton potential in eq.~(\ref{action}). Using Einstein's equations, it is straightforward to show that eq.~(\ref{IR}) follows from a potential of the form:
%%%%%%%%%%%%%%%%%%%%%%%%%%%%%%%%%%%%%%%
\begin{equation}\label{pot}
\mathcal{V}(\f) \to \mbox{const} \times \f^{\frac{\alpha-1}{\alpha}}e^{2\sqrt{\frac{\xi}{D-1}}\f} + \cdots
\end{equation}
%%%%%%%%%%%%%%%%%%%%%%%%%%%%%%%%%%%%%%%
where $\xi$ is the normalization factor appearing in front of the kinetic term in eq.~(\ref{action}). Therefore, $\alpha$ is a parameter of the action, rather than of the particular solution.

The glueball spectrum of the field theory can be obtained by solving the fluctuation equations of the dilaton and the metric, which can generally be expressed as a Schr\"odinger equation of the form $-\psi''(r) + \VS(r)\psi(r) = m^2 \psi(r)$, where $\psi$ denotes a generic glueball wave-function that corresponds to normalizable fluctuations of the bulk fields.\footnote{The Schr\"odinger potential is obtained from the background geometry by fluctuating the bulk fields on the given background and performing a field redefinition to attain the Schr\"odinger form. For example, in the case of spin-$2$ glueballs, it is simply determined by the scale factor of the metric as: $\VS(r) = (D-1) \log''(b_0)/2 + [(D-1)\log'(b_0)/2]^{2}$~\cite{GKN}. In subsection~6.2 of ref.~\cite{GKN}, it is also shown that the spectrum is bounded from below, as the Hamiltonian that appears in the Schr\"odinger problem can be proven to be positive (semi-)definite.}   

Here the Schr\"odinger potential $\VS$ has the following asymptotics:
%%%%%%%%%%%%%%%%%%%%%%%%%%%%%%%%%%%%%%%
\begin{equation}\label{Vs}
\VS(r) \propto r^{-2}\,\,\,\, \mbox{for}\,\,\, r\to 0; \qquad \VS(r) \propto r^{\alpha}\,\,\,\, \mbox{for}\,\,\, r\to \infty.
\end{equation}
%%%%%%%%%%%%%%%%%%%%%%%%%%%%%%%%%%%%%%%
Therefore we observe that the Schr\"odinger potential is bounded both in the ultraviolet (UV) and in the IR limits, hence the glueball spectrum is gapped and discrete, if and only if $\alpha>1$~\cite{GKN}. We will take $\alpha>1$ in the rest of our discussion. In particular, in the limit of large mass, one can use the WKB approximation to write down an approximate expression for the glueball spectra:
%%%%%%%%%%%%%%%%%%%%%%%%%%%%%%%%%%%%%%%
\begin{equation}\label{WKB}
m_n^2 \to C n^{\alpha-1}, \qquad \mbox{for}\,\,\, n\gg 1,
\end{equation}
%%%%%%%%%%%%%%%%%%%%%%%%%%%%%%%%%%%%%%%
where $C$ is a constant that depends on the particular model and on the type of glueball.

\paragraph{Thermodynamics:} The temperature is introduced by Wick-rotating the time direction $t\to i\tau$ and compactifying the Euclidean time: $\tau \sim \tau +1/T$.
%%%%%%%%%%%%%%%%%%%%%%%%%%%%%%%%%%%%%%%
\begin{equation}\label{TG}
  ds^2 = b^2_0(r)\lef( dr^2 + dx_{D-1}^2 + d\tau^2 \ri), \qquad \f=
  \f_0(r).
\end{equation}
%%%%%%%%%%%%%%%%%%%%%%%%%%%%%%%%%%%%%%%
This geometry corresponds to the thermal ensemble in the {\em confined phase}, that we call the {\em thermal gas} (TG) solution. When evaluated on the TG solution, the action in eq.~(\ref{action}) yields the free energy of the dual theory. Neglecting the action counter-terms---which are denoted by the ellipsis in eq.~(\ref{action})---, the result turns out to be divergent, due to the infinite volume of the asymptotic AdS space. Here we shall adopt a particular choice of renormalization, which corresponds to tuning the counter-terms, so that the on-shell thermal gas action vanishes.\footnote{Note that this does not mean that the gas of glueballs in the confined phase has trivial thermodynamics: the non-trivial behavior will be encoded in the determinant of bulk fluctuations around this solution, hence it will be suppressed by $1/N^2$ with respect to the classical saddle-point solution (that, for the TG, is set to zero by our choice of the counter-term action).}

The deconfined phase of the field theory is described by another solution with the same UV asymptotics, the {\em black-hole} (BH) background:
%%%%%%%%%%%%%%%%%%%%%%%%%%%%%%
\begin{equation}\label{BH}
  ds^2 = b^2(r)\left[ f^{-1}(r)dr^2 + dx_{D-1}^2 + d\tau^2 f(r) \right], \qquad \f=
  \f(r).
\end{equation}
%%%%%%%%%%%%%%%%%%%%%%%%%%%%%%
The scale factor $b(r)$ and the dilaton $\f(r)$ are generally different from their counterparts in the thermal gas, eq.~(\ref{TG}).
The blackness function $f(r)$ can be solved in terms of $b(r)$ using Einstein's equations as:
%%%%%%%%%%%%%%%%%%%%%%%%%%%%%%
\be
f(r)=1 - \frac{\int^{r}_{0} dr' b(r')^{1-D}}{\int_{0}^{r_h} dr' b(r')^{1-D}}.
 \label{f}
\ee
%%%%%%%%%%%%%%%%%%%%%%%%%%%%%%
It is a monotonically decreasing function starting as $f=1$ at $r=0$ and vanishing at $r=r_h$. The latter corresponds to the event horizon of the black-hole, where the $00$-component of the BH metric in eq.~(\ref{BH}) vanishes. The temperature of the deconfined gluonic ensemble is given by the Hawking temperature of the BH:
%%%%%%%%%%%%%%%%%%%%%%%%%%%%%%
\be\lab{T}
T = -\frac{f'(r_h)}{4\pi}.
\ee
%%%%%%%%%%%%%%%%%%%%%%%%%%%%%%
The location of the horizon $r_h$ determines the temperature of the system.
The {\em entropy density} of the ensemble is given by the Bekenstein-Hawking entropy:
%%%%%%%%%%%%%%%%%%%%%%%%%%%%%%
\be\lab{s}
{\bf s}(r_h) = \frac{S}{V_{D-1} N^2} = 4\pi \left[ \MP b(r_h) \right]^{D-1},
\ee
%%%%%%%%%%%%%%%%%%%%%%%%%%%%%%
where we defined the entropy density dividing the total entropy by the total spatial volume and by the square of the number of colors.\footnote{More precisely, it should be divided by $N^2-1$, but we work in the large-$N$ limit where the difference becomes irrelevant.} In order to obtain ${\bf s}$ as a function of $T$, one has to invert the variables using eq.~(\ref{T}).

The {\em free energy} of the deconfined phase is given by the value of the action in eq.~(\ref{action}), evaluated on the BH background. As we have fixed the counter-terms for the gravity action in eq.~(\ref{action}) by the requirement that they cancel the on-shell value of the TG action $\cA(TG)$, the free energy is given by $F = \cA(BH) - \cA(TG)$, hence the terms denoted by the ellipsis in eq.~(\ref{action}) can be ignored, because they cancel in the difference.

A practical way to determine the thermodynamics of the gluon plasma is as follows. The general relativistic system satisfies the first law of thermodynamics (see ref.~\cite{GKMN2} and references therein), therefore one can obtain the free energy density directly by integrating the entropy,
%%%%%%%%%%%%%%%%%%%%%%%%%%%%%%
\be\lab{free}
{\bf f}(r_h) = \frac{F}{V_{D-1} N^2} = \int_{r_h}^\infty {\bf s}(r'_h) \frac{d T(r'_h)}{d r'_h} dr'_h .
\ee
%%%%%%%%%%%%%%%%%%%%%%%%%%%%%%
Here the functions ${\bf s}(r_h)$ and $T(r_h)$ are defined in eq.~(\ref{s}) and in eq.~(\ref{T}). The reason for the upper bound of integration is as follows. As the location of the horizon $r_h$ tends to infinity, the size of the BH becomes smaller and smaller and for $r_h \to \infty$ the BH and TG solutions coincide~\cite{GKMN2}. Therefore, the difference between the actions of the two solutions, hence the free energy $F$, should vanish. One can also notice this by comparing the metric functions $b(r)$ and $b_0(r)$ of the BH and of the TG solutions in the limit $r_h\to\infty$, and observe that the difference $b(r)-b_0(r)$ vanishes exponentially in $r_h$~\cite{GKMN2,G1}, in the entire region $0\leq r < r_h$.

Once the free energy is determined as in eq.~(\ref{free}), one can change the variable to $T$ instead of $r_h$, using eq.~(\ref{T}), and obtain the free energy $F(T)$. Given $F(T)$, all other thermodynamic variables follow by standard thermodynamic identities. In particular, one can demonstrate~\cite{GKMN2} the existence of a {\em Hawking-Page phase transition} at a finite temperature $T_c$, if the parameter $\alpha$ in eq.~(\ref{IR}) satisfies the condition: $\alpha\geq 1$~\cite{GKMN1}. The Hawking-Page transition corresponds to the confinement-deconfinement transition in the dual field theory.

We are particularly interested in the behavior of the interaction measure $\Delta$ as a function of $T$. We define the normalized interaction measure by:
%%%%%%%%%%%%%%%%%%%%%%%%%%%%%%
\be\lab{intmes}
\ct(T) = \frac{\Delta}{N^2 T^D} = D \frac{\cf}{T^D} + \frac{{\bf s}}{T^{D-1}},
\ee
%%%%%%%%%%%%%%%%%%%%%%%%%%%%%%
where we used the standard thermodynamic relation: $\Delta = (DF + ST)/V_{D-1}$. The result can be immediately obtained, once the black-hole geometry is found, using the formulas given above. This is carried out explicitly in the next subsection.

\subsection{Model construction}

In this subsection, we shall discuss a general holographic construction that describes a normalized interaction measure $\ct$ which decays with the temperature as $1/T$ in $D=2+1$ dimensions. A more ambitious aim would be to construct a holographic model that fits all lattice data presented in section~\ref{sec:results}. This would certainly be possible by engineering the dilaton potential and fixing the parameters of the model, which are the parameters in the dilaton potential, the integration constants of the equations of motion for $b(r)$ and $\f(r)$, and $\MP$. This aim was successfully achieved in the case of $D=3+1$~\cite{GKMN1,GKMN3,Panero:2009tv}. Here, however, we restrict our investigation to the general behavior of the interaction measure and leave a more detailed holographic construction to future work.

For this purpose, one can use the following simple, semi-analytic construction. Instead of starting from a given dilaton potential and obtaining the metric functions $b(r)$ and $f(r)$ by solving the Einstein's equations, one can simply make an \emph{Ansatz} for the scale factor $b(r)$, obtain $f(r)$ from eq.~(\ref{f}), and derive the thermodynamic functions using the formulas presented in the previous subsection. Although this method is {\em not exact} but only approximate,\footnote{In particular, the solution that one finds in this manner does not solve Einstein's equations for a given dilaton potential $V(\f)$, but rather determines the potential. This potential will generally depend on $r_h$ beyond a certain value of $\phi$, where the approximation breaks down~\cite{Kajantie}.} it turns out to provide a good approximation to thermodynamics in a large range of temperatures, as has been discussed in ref.~\cite{Kajantie}. The general conclusion, that we derive in the following, will be independent of the details of this approximation.

At this point, the question is: What \emph{Ansatz} should one take for the scale factor $b(r)$? To answer this question, we first recall that $b(r)$ tends to its TG analogue $b_0(r)$ both in the UV and in the IR limits, therefore eq.~(\ref{IR}) leads to:
%%%%%%%%%%%%%%%%%%%%%%%%%%%%%%
\bea\lab{lim1}
b(r)&\to& e^{-(r\La)^\alpha + \cdots}, \qquad \mbox{as}\,\,\, r\to\infty\\
b(r)&\to& \frac{\ell}{r} + \cdots \qquad \mbox{as}\,\,\, r\to 0,
\lab{lim2}
\eea
%%%%%%%%%%%%%%%%%%%%%%%%%%%%%%
where the ellipsis denotes subleading terms. A simple \emph{Ansatz} that satisfies both limits is:\footnote{Although this is a well-motivated \emph{Ansatz}, clearly it is not the most general behavior for the scale factor. In particular, in the intermediate $r$ region, a less restrictive \emph{Ansatz}, parametrized by variables in addition to $\alpha$, would allow for a better fit of the lattice data, especially near $T_c$. However, as emphasized at the beginning of this subsection, such a general construction is not our primary purpose in this paper. }
%%%%%%%%%%%%%%%%%%%%%%%%%%%%%%
\be\lab{b}
b(r) = \frac{\ell}{r} e^{-(r\La)^\alpha}, \qquad \mbox{with}\,\,\, \alpha>1.
\ee
%%%%%%%%%%%%%%%%%%%%%%%%%%%%%%
As discussed after eq.~(\ref{IR}), the parameter $\alpha$ controls the properties of the theory in the IR. In particular, $\alpha$ should be larger than $1$ for a confining theory. Interestingly, $\alpha$ also controls the nature of the confinement-deconfinement phase transition at $T_c$: {\em The transition tends to become a continuous one (in particular, a second-order one) as $\alpha \to 1$}~\cite{G1,G2}.
%%%%%%%%%%%%%%%%%%%%%%%%%%%%%
\begin{figure}[ht]
\begin{center}
\includegraphics[width=.75\textwidth]{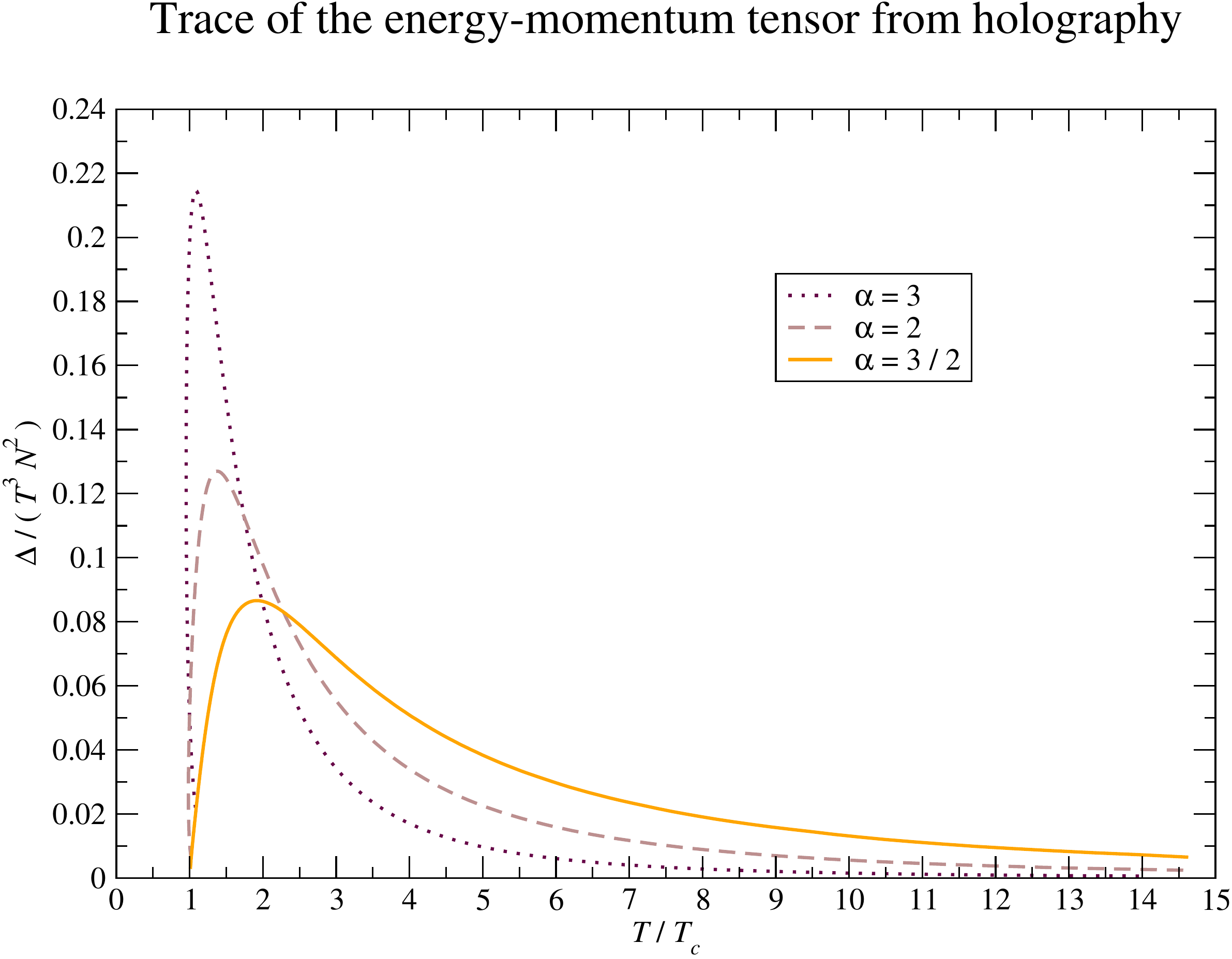}
\caption{Comparison of the interaction measure $\Delta$ (normalized by $T^3$ and by $N^2$) that follows from the holographic construction, for different choices of the parameter $\alpha$. The solid (orange) curve is obtained for $\alpha=3/2$, the dashed (brown) curve corresponds to $\alpha=2$, and finally the dotted (maroon) curve is the result for $\alpha=3$.}
\lab{fig:various_alpha}
\end{center}
\end{figure}%
%%%%%%%%%%%%%%%%%%%%%%%%%%%%%
Here we show that the parameter $\alpha$ also controls the decay of the interaction measure in the range of temperatures between $T_c$ and some intermediate value $T_i \gg T_c$. The interaction measure is obtained from eq.~(\ref{intmes}) by numerical integration.\footnote{The upper integration limit in eq.~(\ref{intmes}) can be chosen to be a large enough number, so that the change in the integral is numerically negligible if the integration limit is pushed to larger values.} Fig.~\ref{fig:various_alpha} shows the interaction measure $\Delta$ (normalized dividing by $T^3 N^2$) as a function of the temperature $T$ (in units of the deconfinement temperature $T_c$), for different values of $\alpha$, in the case of $D=2+1$ spacetime dimensions. We observe that the curve becomes steeper with increasing values of $\alpha$; in section~\ref{sec:results}, we show a comparison of the curve corresponding to $\alpha = 3/2$ to the numerical results obtained from lattice simulations, see fig.~\ref{fig:nt6_rescaled_trace}. 
One technicality in our present calculation is the value of $T_c$. In general it is a number of the same order as $\La$:
%%%%%%%%%%%%%%%%%%%%%%%%%%%%%%
\be\lab{TL}
T_c = c_0 \La,
\ee
%%%%%%%%%%%%%%%%%%%%%%%%%%%%%%
with $c_0$ being some constant depending on the model. In order to determine $c_0$ (for a given $\La$), we search for a value $r_c$, such that for $r_h=r_c$ the free energy difference vanishes: $\cf(r_c)=0$. By definition, this point corresponds to the phase transition. The transition temperature $T_c$ is then obtained by evaluating eq.~(\ref{T}) at $r_h=r_c$.

\subsection{General conclusions from holography}

One can understand the dependence of the slope of the reduced interaction measure on the parameter $\alpha$ semi-analytically as follows. From eq.~(\ref{intmes}), one can write:
%%%%%%%%%%%%%%%%%%%%%%%%%%%%%%
\be\lab{intmes1}
\ct(T) = \frac{\Delta}{N^2 T^D} = D \frac{\cf}{T^D} + \frac{{\bf s}}{T^{D-1}}
= \frac{{\bf s}}{T^{D-1}} - \frac{D}{T^D} \int_{T_c}^T {\bf s}(\tT) d\tT,
\ee
%%%%%%%%%%%%%%%%%%%%%%%%%%%%%%
where, again, we used the standard thermodynamic relation $\Delta = (DF + ST)/V_{D-1}$ and we rewrote the free energy $F$ using the first law as
%%%%%%%%%%%%%%%%%%%%%%%%%%%%%%
\be\lab{free1}
F(T) = -\int_{T_c}^T S(\tilde{T}) d\tilde{T}
\ee
%%%%%%%%%%%%%%%%%%%%%%%%%%%%%%
(where the choice for the lower bound of the integral comes from the requirement that $F(T_c)=0$ at the transition). Note that, for extremely high temperatures, the expression on the right-hand side of eq.~(\ref{intmes1}) vanishes, by asymptotic conformality of the theory. Technically, for very large $T$ the leading-order term in a $1/T$ expansion of the integral cancels the first term in eq.~(\ref{intmes1}). The contribution that we seek for is the sub-leading term of the integral in $1/T$.

For large enough temperatures $T/\La\gg 1$, one can use the asymptotically conformal result to relate $T$ to $r_h$. This follows from the AdS black-hole expression where the blackness function in eq.~(\ref{BH}) becomes: $f = 1 - (r_h/r)^D$. As the background turns into the asymptotically AdS BH for large $T$, substituting this expression in eq.~(\ref{T}) one finds:
%%%%%%%%%%%%%%%%%%%%%%%%%%%%%%
\be\lab{Tads}
T  \approx  \frac{D}{4\pi r_h}, \qquad \mbox{for}\,\,\, T \gg \La.
\ee
%%%%%%%%%%%%%%%%%%%%%%%%%%%%%%
Using this expression in eq.~(\ref{b}) and in eq.~(\ref{s}), one immediately obtains an approximate expression from eq.~(\ref{intmes1}) as:
%%%%%%%%%%%%%%%%%%%%%%%%%%%%%%
\be\lab{intmes2}
\ct(T) = c_1 \left[ e^{-c_2 \lef(\frac{T}{T_c}\ri)^{-\alpha}} - \frac{D}{T^D} \int_{T_c}^{T} d\tT ~\tT^{D-1} e^{-c_2 \lef(\frac{\tT}{T_c}\ri)^{-\alpha}} \right],
\ee
%%%%%%%%%%%%%%%%%%%%%%%%%%%%%%
where we defined the constants:
%%%%%%%%%%%%%%%%%%%%%%%%%%%%%%
\be\lab{consts}
c_1 = 4\pi\lef(\frac{4\pi \MP\ell}{D}\ri)^{D-1}, \qquad c_2 = (D-1)\lef(\frac{D}{4\pi c_0}\ri)^\alpha,
\ee
%%%%%%%%%%%%%%%%%%%%%%%%%%%%%%
while $c_0$ is defined in eq.~(\ref{TL}). Evaluating the integral in eq.~(\ref{intmes2}) for large $T$ yields:
%%%%%%%%%%%%%%%%%%%%%%%%%%%%%%
\be\lab{intmes3}
\ct(T) \approx \frac{c_1c_2\alpha}{(D-\alpha)} \lef(\frac{T}{T_c}\ri)^{-\alpha}\left[1 + \cO\lef(\frac{T}{T_c}\ri)^{-\alpha}\right] .
\ee
%%%%%%%%%%%%%%%%%%%%%%%%%%%%%%
Clearly, this expression is valid only for $\alpha\ne D$. The case $\alpha = D$ should be treated separately, and one finds: 
%%%%%%%%%%%%%%%%%%%%%%%%%%%%%%
\be\lab{intmes4}
\ct(T) \approx c_1c_2\;D\log\left(\frac{T}{T_c}\right) \lef(\frac{T}{T_c}\ri)^{-D}\left[1 + \cO\lef(\frac{T}{T_c}\ri)^{-D}\right] .
\ee
%%%%%%%%%%%%%%%%%%%%%%%%%%%%%%
 
The conclusion of this semi-analytic calculation is that, as $T$ increases, the normalized interaction measure falls off as a function of $T/T_c$, and the precise shape of the fall-off is a power-law determined by the parameter $\alpha$. In particular, this power is {\em not} directly related to the spacetime dimensionality $D$, but rather to the nature of the deconfinement transition, which is determined by the value of the exponent $\alpha$ in eq.~(\ref{IR}).

This is in agreement with known results from lattice computations. In particular:
\begin{itemize}
\item In $D=3+1$ dimensions, for pure Yang-Mills theory one expects linear confinement with an asymptotically {\em linear} glueball spectrum. Then, eq.~(\ref{WKB}) determines $\alpha=2$ and from eq.~(\ref{intmes3}) one expects $\ct \sim 1/T^2$ for some range of temperatures above $T_c$. This has indeed been observed in lattice simulations~\cite{Panero:2009tv,SU3_EoS,Tsquare_in_4D}. 
\item In $D=2+1$ dimensions, the confinement-deconfinement transition tends towards being continuous. In particular, for $N=2$ and $3$ it is known to be a second-order transition, and for $N=4$ it may be continuous or a weakly first-order one; a general statement that one can make is that Yang-Mills theories in $2+1$ dimensions are more inclined to exhibit continuous or weakly first-order transitions than in $D=3+1$~\cite{Liddle:2008kk, Bialas:2008rk, SUN_thermodynamics_in_2_plus_1_dimensions}. On the other hand, as we discussed above, the nature of the transition in holography is determined by the exponent $\alpha$. For $D=2+1$ one expects\footnote{In $D=2+1$ dimensions, one is tempted to set $\alpha=1$ for the continuous transitions for $N=2$ and $3$, by the arguments in refs.~\cite{G1,G2}. However, these arguments hold in the large-$N$ limit, and $\alpha$ should receive $1/N$ corrections for finite, and small, values of $N$.} it to be smaller than $2$. Furthermore, in order to have confinement at zero temperature, $\alpha$ should also be larger than $1$, see eq.~(\ref{IR}). Therefore we expect $1<\alpha<2$ for the $D=2+1$ theories under study. Indeed, as discussed in sec.~\ref{sec:results}, we found that $\alpha=3/2$ gives a very good fit to the lattice results.

\end{itemize}

One should be cautious with the various approximations that we made in this section. First of all, we adopted a semi-analytic approach in determining the holographic background, by setting the scale factor of the metric via eq.~(\ref{b}). In the well-studied case of $3+1$ dimensions, this approximation works quite well, as shown in ref.~\cite{Kajantie}. Guided by these results, we made an educated guess for the scale factor of the metric. Another approximation is to treat the relation between $T$ and $r_h$ as in the case of AdS, eq.~(\ref{Tads}). Strictly speaking, this is only valid for large $T$, where $r_h$ is small enough, so that the asymptotically AdS region sets in. However, numerical studies show that this is also a good approximation in a large range of temperatures, ranging all the way from the limit of infinite $T$ down to near $T_c$~\cite{GKMN3, Kajantie}. The final approximation is the large-$N$ limit. Although we believe that the first two approximations can be justified at least in a finite range of temperatures, the validity of the latter cannot be assessed by simple arguments, but should be checked through case-by-case studies. In the case of $3+1$ dimensions, the $\SU(N)$ Yang-Mills lattice data show that, indeed, the equilibrium thermodynamic quantities per gluon are essentially independent of $N$~\cite{Panero:2009tv}, and agree well with holographic calculations~\cite{GKMN3, Gubser:1998nz}. We hope that this general conclusion also holds for theories in $2+1$ dimensions. This should be checked by a thorough study of holographic models in $D=2+1$, that we plan to pursue in the future. In the following sections, after introducing the setup to study $\SU(N)$ Yang-Mills theories in $D=2+1$ dimensions on the lattice, we compare the lattice results for the equation of state with the predictions of the holographic model that we discussed in this section.

\section{Yang-Mills theories in $2+1$ dimensions}
\label{sec:2+1_gauge_theories}

The continuum formulation of Yang-Mills theories with $\SU(N)$ gauge group in $D=2+1$ spacetime dimensions can be defined via the Euclidean functional integral:
\eq
\label{continuum_formulation}
Z = \int\mathcal{D} A e^{-\Se}, \;\;\; \Se = \int {\dd}^3x \frac{1}{2 g_0^2 }\Tr F_{\alpha\beta}^2,
\en
where $g_0^2$ (which has the dimensions of an energy) is the bare square gauge coupling, $F_{\alpha\beta}(x)$ denotes the non-Abelian field strength tensor, and the functional integration is done over the non-Abelian gauge field $A_\mu(x)$, taking values in the adjoint representation of the algebra of the gauge group. Like in $D=3+1$, $\SU(N)$ Yang-Mills theories are also asymptotically free in $D=2+1$ dimensions; since $g_0^2$ is dimensionful, perturbative computations for processes at a momentum scale $k$ can be organized as series in powers of the ratio $g_0^2/k$~\cite{3d_YM_renormalization_properties}.

To make the definition in eq.~(\ref{continuum_formulation}) mathematically well-defined at the non-perturbative level, one can introduce a gauge-invariant lattice regularization. The common choice is a regularization on a cubic, isotropic lattice $\Lambda$ of spacing $a$, which allows one to trade the continuum field $A_\mu(x)$ for an at most countable (and finite, if the spacetime is truncated to a finite volume) set of matrices $ U_\mu(x) $, which are defined on the oriented links joining nearest-neighbor lattice sites. The $ U_\mu(x) $ matrices represent parallel transporters on the lattice links, and take values in the adjoint representation of the gauge group. Their dynamics is defined by:
\eq
\label{lattice_partition_function}
\Zlat = \int \prod_{x \in \Lambda} \prod_{\alpha=1}^3 {\dd}U_\alpha(x) e^{-\Sel},
\en
where ${\dd}U_\alpha(x)$ denotes the Haar measure for each $U_\alpha(x)$, and $\Sel$~is the Wilson gauge action~\cite{Wilson:1974sk}:
\eq
\label{Wilson_lattice_gauge_action}
\Sel= \beta \sum_{x \in \Lambda} \sum_{1 \le \alpha < \beta \le 3} \left[1 - \frac{1}{N} \re\Tr U_{\alpha\beta}(x)\right], \;\;\; \mbox{with:} \;\;\; \beta=\frac{2N}{g_0^2 a},
\en
which tends to $\Se$ in the na\"{\i}ve (i.e. tree-level) continuum limit $a \to 0$, with corrections $\mathcal{O}(a^2)$, and which, being defined in terms of the trace of the plaquette variable:
\eq
\label{plaquette}
U_{\alpha\beta}(x) = U_\alpha(x) U_\beta(x+a\hat\alpha) U^\dagger_\alpha(x+a\hat\beta) U^\dagger_\beta(x),
\en
is exactly gauge-invariant at all values of the lattice spacing $a$.

The expectation value of a physical observable $O$ on the lattice is defined by:
\eq
\label{expectation_value}
\langle O \rangle = \frac{1}{\Zlat} \int \prod_{x \in \Lambda} \prod_{\alpha=1}^3 {\dd}U_\alpha(x) \; O \; e^{-\Sel}.
\en
This quantity is a ratio of high-, but finite-dimensional, finite, ordinary group integrals, and can be estimated numerically by importance sampling over an ensemble of configurations of link matrices.

Previous lattice computations~\cite{Liddle:2008kk,  Bialas:2008rk, SUN_thermodynamics_in_2_plus_1_dimensions} have shown that, similarly to the $D=3+1$ case, also $D=2+1$ non-Abelian gauge theories exhibit linear confinement at low energy (i.e. the interquark potential $V(r)$ grows as $ \sigma r$ at large distances $r$) and a gapped, discrete spectrum of color-singlet glueball states. Furthermore, they also undergo a deconfining phase transition at a finite temperature $T_c$, associated with the spontaneous breakdown of a global center symmetry.

In order to ``set the scale'' (i.e., to determine the value of the spacing $a$ as a function of $\beta$) we used the accurate non-perturbative results reported in ref.~\cite{Liddle:2008kk}, which lead to the following expression for the temperature (in units of $T_c$) as a function of $\beta$:
\eq
\label{scale_setting}
\frac{T}{T_c} = \frac{\beta - 0.22 N^2 + 0.5}{N_t \cdot \left( 0.357 N^2 + 0.13 - 0.211/N^2 \right)} \;.
\en
The statistical and systematic uncertainties in this scale determination are set by those on the value of the critical temperature over the square root of the zero-temperature string tension $\sigma$ in the continuum limit and in the large-$N$ limit from ref.~\cite{Liddle:2008kk}. The latter reports $T_c/\sqrt{\sigma}=0.9026(23)$ in the $N \to \infty$ limit of this ratio, with a finite-$N$ correction term proportional to $N^{-2}$ (and valid down to $N=2$), whose coefficient is $0.880(43)$. So our uncertainty on the temperature scale can be estimated to be of the order of $1\%$, and does not have a real impact on our analysis (hence, for the sake of clarity, in our plots we do not show the errorbars on the temperature). Another potential source of systematic uncertainties is given by the choice of the physical observable to set the scale: since all numerical simulations are performed at finite values of the spacing $a$, the determination of the physical scale using different observables can be affected by different discretization artifacts; however, the quantitative effect of the induced systematic uncertainty is small, $\mathcal{O}(a^2)$. For a comparison with alternative non-perturbative definitions of the scale, see, e.g., refs.~\cite{Bialas:2008rk, SUN_thermodynamics_in_2_plus_1_dimensions, Caselle:2004er}. Finally, note that, in principle, one could also use a perturbative definition of the scale; in particular, for the $D=3+1$ case high-order perturbative computations in the lattice scheme are available in the literature~\cite{lattice_perturbation_theory}. However, since in the present work we are interested in a temperature regime where non-perturbative effects are expected to be non-negligible, our determination of the scale is completely non-perturbative.

In our simulations, we generated the gauge configurations using code implementing  a $3+1$ combination of local overrelaxation and heat-bath updates on $\SU(2)$ subgroups~\cite{algorithm}; for part of our simulations, we also used the Chroma suite~\cite{Edwards:2004sx}. In the following, we denote the cardinality of our configuration ensembles as $\nconf$. Having defined the lattice action via eq.~(\ref{Wilson_lattice_gauge_action}), the only parameters that fix the physical setup of our simulations are the number of colors $N$, the Wilson gauge action parameter $\beta$, and the sizes of the lattice along the space-like and time-like directions, which can be expressed in units of the lattice spacing $a$ as $aN_s$ and $aN_t$, respectively. As usual in a Euclidean QFT setup (assuming periodic boundary conditions for the bosonic fields), the latter quantity is related to the physical temperature via $aN_t=T^{-1}$.

The parameters of the simulations that we performed for this work are summarized in table~\ref{tab:parameters}; we chose $N_s \gg N_t$, which guarantees a good approximation of the thermodynamic limit~\cite{Gliozzi_finite_volume}. 
%%%%%%%%%%%
In fact, finite-volume effects in the $T > T_c$ phase are known to be strongly suppressed, due to the screening phenomenon in the deconfined plasma. Note that, as pointed out in ref.~\cite{Gliozzi_finite_volume} (for the $D=3+1$ case), the thermodynamics of a gas of free gluons \emph{is} sensitive to finite-volume corrections, which depend on the product of the linear spatial size of the lattice times the temperature. In the limit of very high temperatures, such corrections lead to quantifiable corrections to ordinary thermodynamic relations. However, at the relatively moderate temperatures probed in the present lattice simulations, the numerical evidence from all previous studies (both in $D=3+1$ and in $D=2+1$ dimensions) indicates that screening makes finite-volume corrections essentially negligible for simulations on lattices with $N_s/N_t \ge 4$. In particular, the accurate numerical study of $\SU(3)$ thermodynamics in $D=2+1$ dimensions presented in ref.~\cite{Bialas:2008rk} provided convincing evidence for the strong suppression of finite-volume effects. On the other hand, deviations from the thermodynamic limit in the confining phase are exponentially suppressed by the finiteness of the mass gap: if $m_{0}$ denotes the mass of the lightest glueball, then, typically, lattices of linear size $aN_s \ge 4/m_{0}$ are such, that systematic effects due to the volume finiteness play an essentially negligible r\^ole in the lattice computation error budget (which is dominated by finite-cutoff effects, and by statistical uncertainties due to the finite cardinality of the sampled configuration ensemble).
%%%%%%%%%%%
The results at zero temperature are obtained from simulations on cubic lattices of volume $(a N_s)^3$. In addition to the simulations listed in table~\ref{tab:parameters}, for the $\SU(2)$ and $\SU(4)$ gauge groups we also analyzed the configurations corresponding to the $N_t=6$ and $N_t=8$ ensembles (and their respective $T=0$ counterparts) taken from ref.~\cite{Caselle:2011fy}.

The equation of state of Yang-Mills theories in $D=2+1$ dimensions can be easily obtained from elementary thermodynamic identities. Let $Z(T,V)$ denote the partition function for an isotropic system of two-dimensional ``volume'' $V$ at temperature $T$; in the thermodynamic limit $V \to \infty$, the pressure $p$ is related to the free energy $F=-T\ln Z$ by $pV=-F$, and to the trace of the energy-momentum tensor $\Delta=T^\mu_{\phantom{\mu}\mu}$ via:
\begin{equation}
\label{delta_and_p}
\frac{\Delta}{T^3} = T \frac{d}{d T} \left(\frac{p}{T^3}\right).
\end{equation}
As discussed above, deviations from the thermodynamic-limit relation $pV=-F$ due to the finiteness of the lattice volume can be neglected at the temperatures investigated in this work. Finally, the energy and entropy densities (denoted as $\epsilon$ and $s$, respectively) can be obtained from $\epsilon =  \Delta + 2p$ and $ sT = \Delta + 3p$.

Our determination of the equation of state on the lattice is done according to the ``integral method''~\cite{Engels:1990vr}: the trace of the energy-momentum tensor is extracted from differences of $\langle U_\Box \rangle_{T}$, the expectation value of the average trace of the plaquette at a temperature $T$:
\eq
\label{Delta_lattice}
\Delta  = \frac{3}{a^3} \frac{\partial \beta}{\partial \ln a} \left( \langle U_\Box \rangle_{T} - \langle U_\Box \rangle_{0} \right),
\en
so that the pressure is obtained by integration over $\beta$:
\eq
\label{p_lattice}
p  = \frac{3}{a^3} \int_{\beta_0}^\beta \dd \beta^\prime \left( \langle U_\Box \rangle_{T} - \langle U_\Box \rangle_{0} \right),
\en
starting from a lower integration extremum $\beta_0$ corresponding to a temperature sufficiently deep in the confined phase. We performed the numerical evaluation of the integral in eq.~(\ref{p_lattice}) comparing the trapezoid rule with the method described by eq.~(A.4) in ref.~\cite{Caselle:2007yc}, which is characterized by systematic errors $\mathcal{O}(n_\beta^{-4})$. Since our scan in $\beta$ values is very fine, the systematic error related to the choice of the numerical integration method has a negligible r\^ole in the error budget.

\begin{table}
\begin{center}
\begin{tabular}{| c | c | c | c | c | c |}
\hline
$N$  & $N_s^2 \times N_t$ & $n_\beta$ & $\beta$-range
& $\nconf$ at $T=0$ & $\nconf$ at finite $T$ \\
\hline \hline
$2$ & $48^3$          &  $218$ & $[7.2,65.0]$ & $2 \times 10^5 $ & ---  \\
    & $90^2 \times 6$ &       &                & --- & $5 \times 10^5 $  \\
    & $64^3$          &  $157$ & $[10.5,90.0]$  & $2 \times 10^5 $ & ---  \\
   & $120^2 \times 8$ &       &                & --- & $5 \times 10^5 $  \\ \hline
$3$ & $48^2 \times 6$ &  $101$ & $[16.4,146.4]$ & $1 \times 10^5 $ & $8 \times 10^5$  \\
    & $64^2 \times 8$ &  $127$ & $[15.0,189.6]$ & $5 \times 10^4 $ & $4 \times 10^5$  \\ \hline
$4$ & $48^2 \times 6$ &  $261$ & $[30.0,246.0]$ & $2 \times 10^4 $ & $1.6 \times 10^5$  \\
    & $64^2 \times 8$ &  $336$ & $[39.0,324.9]$ & $1.5 \times 10^4$& $1.2 \times 10^5$  \\ \hline
$5$ & $48^2 \times 6$ &  $150$ & $[43.5,386.5]$ & $1 \times 10^5 $ & $8 \times 10^5 $ \\
    & $64^2 \times 8$ &   $46$ & $[60.0,510.0]$ & $1 \times 10^4 $ & $8 \times 10^4 $ \\ \hline
$6$ & $48^2 \times 6$ &  $129$ & $[66.0,561.0]$ & $2.5 \times 10^4 $ & $2 \times 10^5 $ \\ \hline
\end{tabular}
\end{center}
\caption{Parameters of the new lattice simulations performed for this work: $N$ denotes number of colors, $N_t$ and $N_s$ are, respectively, the lattice sizes along the time-like and space-like directions (in units of the lattice spacing). $n_\beta$ denotes the number of $\beta$-values (i.e. of temperatures) that were simulated, in each $\beta_{min} \le \beta \le \beta_{max}$ interval; the $T=0$ and finite-$T$ statistics at each $\beta$-value are shown in the last two columns. For $N>2$, all $T=0$ simulations were performed on lattices of size $(aN_s)^3$. Our analysis also includes part of the data from the simulations reported in ref.~\protect\cite{Caselle:2011fy}.
}
\label{tab:parameters}
\end{table}

\section{Numerical results}
\label{sec:results}

In this section, we present our numerical results for the basic equilibrium thermodynamic properties in $D=2+1$ $\SU(N)$ Yang-Mills theories with $N=2$, $3$, $4$, $5$ and $6$ colors, and compare them to the predictions from the holographic model introduced in section~\ref{sec:holographic_model}. By virtue of asymptotic freedom, one expects that in the high-temperature limit the thermodynamics of these theories reduces to that of a gas of non-interacting gluons, whose equation of state in the continuum reads:
\begin{equation}
\label{free_gluon_gas}
\frac{p}{T^3}=(N^2-1)\frac{\zeta(3)}{2\pi},
\end{equation}
where $\zeta(3) \simeq 1.20205690316\dots$ is Ap\'ery's constant. On the lattice, eq.~(\ref{free_gluon_gas}) is affected by cutoff corrections:
\begin{equation}
\label{lattice_free_gluon_gas}
\frac{p_{\tiny\mbox{L}}}{T^3}=(N^2-1)\frac{\zeta(3)}{2\pi} \tilde{R}_I(N_t)\;,
\end{equation}
where the correction factor $\tilde{R}_I(N_t)$ can be either estimated numerically or evaluated analytically, order by order in an expansion in powers of $N_t^{-2}$~\cite{Engels:1999tk}. For the Wilson action and the integral method that we used in this work, the latter computation yields:
\eq
\tilde{R}_I(N_t) = 1 + \frac{7}{4} \frac{1}{N_t^2} \frac{\zeta(5)}{\zeta(3)} + \frac{227}{32} \frac{1}{N_t^4} \frac{\zeta(7)}{\zeta(3)} + \frac{8549}{128} \frac{1}{N_t^6} \frac{\zeta(9)}{\zeta(3)} + \mathcal{O}\left( ( \pi / N_t )^8 \right)
\label{RI_tilde}
\en
(see the appendix~\ref{app:lattice_SB} for details). However, it is important to stress that the cutoff artifacts encoded by this correction factor are suppressed at temperatures close to $T_c$, and hence we \emph{do not} rescale our numerical results by $\tilde{R}_I(N_t)$. 

Since the right-hand side of eq.~(\ref{free_gluon_gas}) is proportional to the number of gluon degrees of freedom (one transverse polarization state for each of the $N^2-1$ color d.o.f.), in the deconfined phase it is natural to normalize the dimensionless ratios $p/T^3$, $\Delta/T^3$, $\epsilon/T^3$ and $s/T^2$ obtained in gauge theories with a different number of colors, by dividing them by $N^2-1$. Note, however, that, while this is expected to make the results from different groups collapse onto the same curve in the limit of very high temperatures, in principle there is no obvious reason to expect the same to be true also at moderate temperatures close to $T_c$, where the deconfined plasma is far from being weakly coupled (and the thermodynamics could perhaps be dominated by different degrees of freedom, with unknown scaling properties with $N$).

\begin{figure*}
\centerline{\includegraphics[width=.75\textwidth]{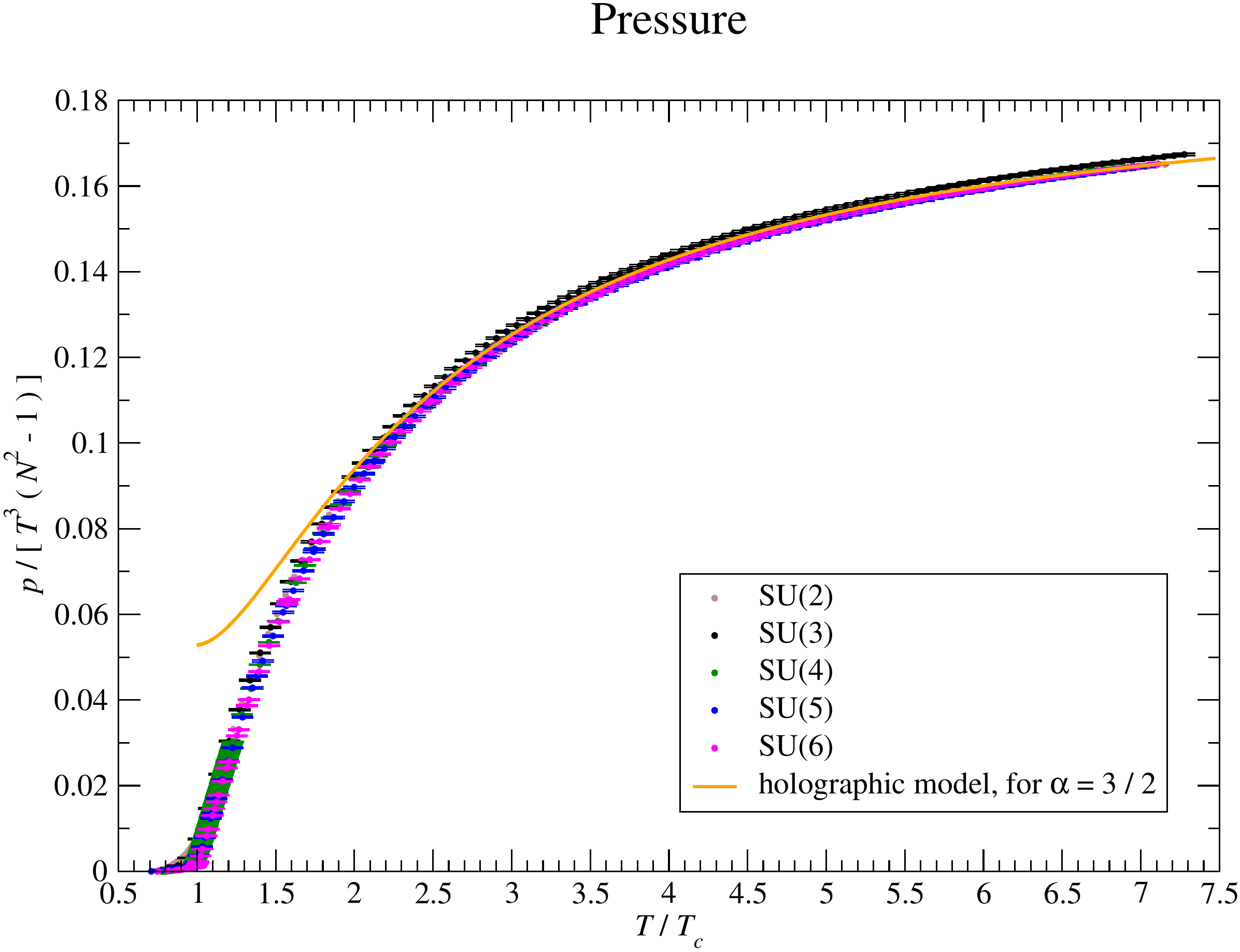}}
\vspace{1cm}
\caption{The pressure per gluon degree of freedom, in units of $T^3$, as a function of the temperature (in units of $T_c$), for the gauge groups $\SU(2)$ (brown), $\SU(3)$ (black), $\SU(4)$ (green), $\SU(5)$ (blue) and $\SU(6)$ (magenta). The plot shows the results obtained from simulations on lattices with $N_t=6$. The solid orange curve is the corresponding prediction from the holographic model discussed in section~\protect\ref{sec:holographic_model}, for $\alpha=3/2$, as obtained by numerical integration of $\Delta/[T^4(N^2-1)]$ over $T$, with an integration constant fixed by imposing consistency of the holographic model with the lattice data at high temperatures.}
\label{fig:nt6_pressure}
\end{figure*}

Figure~\ref{fig:nt6_pressure} shows our results for the pressure per gluon, in units of $T^3$, as a function of $T/T_c$. The plot shows the results that we obtained for the various gauge groups, from simulations on lattices with $N_t=6$ sites in the compactified Euclidean time direction, and for the corresponding space-like volumes listed in table~\ref{tab:parameters}. As we shall discuss in the following, simulating finite-temperature lattices at this value of $N_t$ turns out to be an optimal choice, given that it allows one to reach very precise numerical results for the plaquette differences appearing in eq.~(\ref{p_lattice}), while keeping the systematic effects due to the finiteness of the lattice cutoff under control. The first, striking result manifest from fig.~\ref{fig:nt6_pressure} is the nearly perfect scaling of $p$ with $N^2-1$: in the deconfined phase, the numerical results of the pressure per gluon collapse on the same curve, for all the gauge groups that we simulated (denoted by symbols of different colors). This is analogous to what happens in $3+1$ dimensions~\cite{finiteTlargeNlatticeresults,Panero:2009tv}, and can be clearly contrasted to the behavior in the confining phase, where, on the contrary, all bulk thermodynamic quantities scale proportionally to $\mathcal{O}(N^0)$, i.e. are independent of $N$, in the large-$N$ limit~\cite{Caselle:2011fy}. While in principle it is reasonable to expect that the equation of state should be qualitatively similar in all of these theories, the remarkable quantitative agreement among different gauge groups that our results reveal is completely non-trivial. Generally, for $\SU(N)$ gauge theories without quark fields, the leading-order finite-$N$ corrections with respect to the large-$N$ limit are expected to be proportional to $N^{-2}$, and hence could amount to relative deviations of the order of $10\%$ for $\SU(3)$, or even larger for $\SU(2)$. On the contrary, our high-precision lattice results do not reveal any statistically significant evidence\footnote{The only differences among results corresponding to different gauge groups can be interpreted as statistical fluctuations, and/or in terms of the small systematic uncertainty related to the scale setting, as discussed in section~\protect\ref{sec:2+1_gauge_theories} (for the sake of clarity, the horizontal errorbars associated with the accuracy limits on our temperature determination are not displayed in the figures).} of such dependence on $N$. Figure~\ref{fig:nt6_pressure} also reveals that the approach to the continuum Stefan-Boltzmann limit (which is about $0.1913\dots$, out of the vertical axis range) is relatively slow: at temperatures slightly above $7~T_c$, the pressure is still approximately $15\%$ off from the Stefan-Boltzmann value, indicating that the plasma is still far from an ideal gas of free massless gluons. A qualitatively similar feature is also observed for Yang-Mills theories in $3+1$ dimensions~\cite{Panero:2009tv,SU3_EoS} (for which, however, it is important to observe that the energy scale dependence of the physical coupling is different). Finally, figure~\ref{fig:nt6_pressure} also shows the prediction (solid orange curve) from the holographic model discussed in section~\ref{sec:holographic_model}, for $\alpha=3/2$. Since the holographic prediction for $p/[T^3(N^2-1)]$ is obtained by integration of $\Delta/[T^4(N^2-1)]$ over $T$, based on eq.~(\ref{delta_and_p}), we fixed the integration constant by imposing consistency of the holographic model with the lattice data at high temperatures.\footnote{Note that the approximations involved in the construction of the simple holographic model considered here lead to some deviations from the lattice data for temperatures close to the transition region, and, in particular, to an unphysical non-vanishing value for $p/[T^3(N^2-1)]$ for $T \to T_c^+$. Since the pressure is a continuous function of the temperature, this would imply a non-vanishing value for $p/[T^3(N^2-1)]$ also for $T \to T_c^-$, in clear contradiction with the fact that the number of physical degrees of freedom in the confining phase scales like $\mathcal{O}(N^0)$, not like $\mathcal{O}(N^2)$, at large $N$.}

\begin{figure*}
\centerline{\includegraphics[width=.75\textwidth]{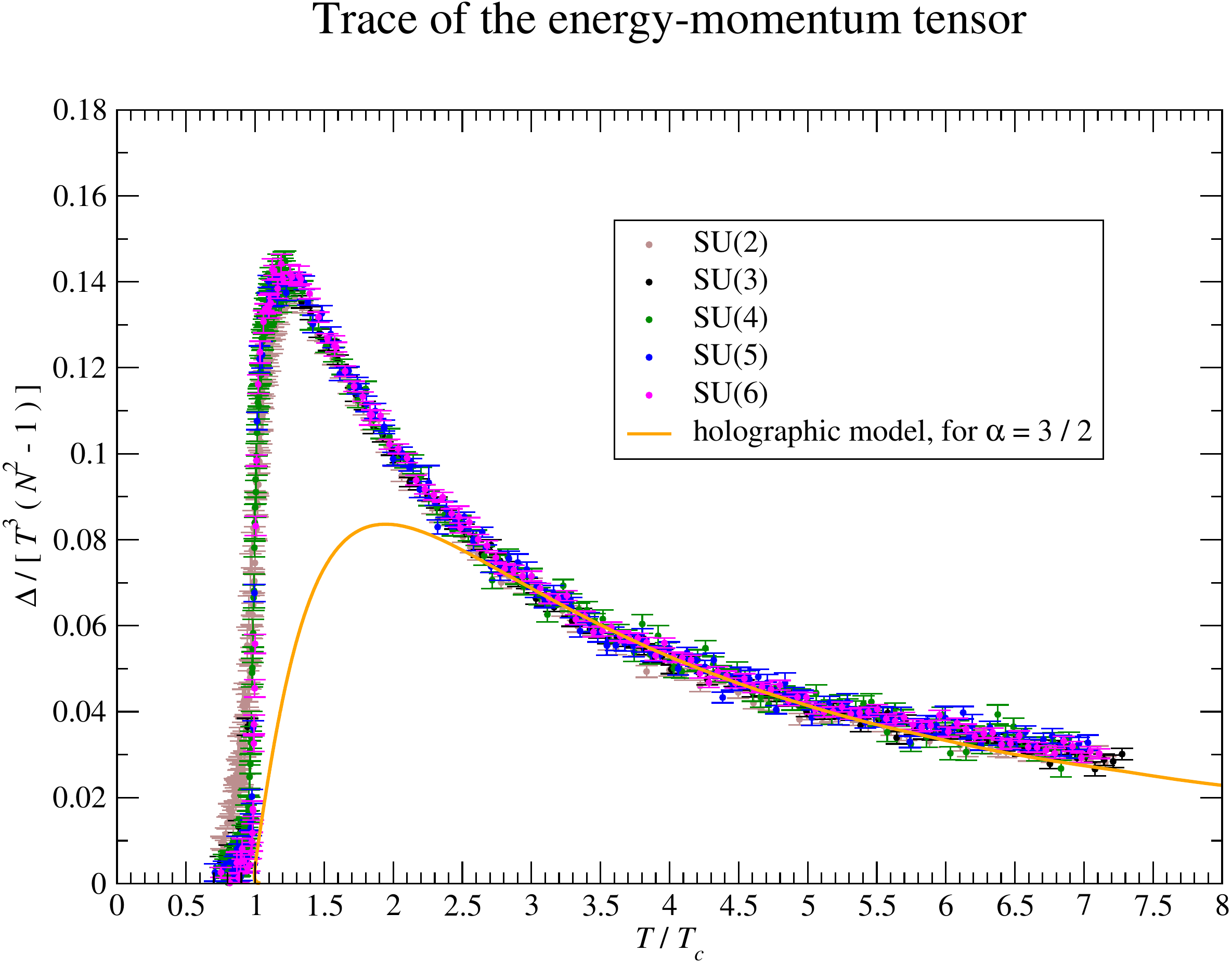}}
\vspace{1cm}
\caption{Same as in fig.~\ref{fig:nt6_pressure}, but for the trace of the energy-momentum tensor per gluon, in units of $T^3$. The solid orange curve is the corresponding prediction from the holographic model discussed in section~\protect\ref{sec:holographic_model}, for $\alpha=3/2$. Note that, in principle, the holographic model could be refined, to match the lattice data also at temperatures lower than $2.5~T_c$, through an appropriate choice of the dilaton potential. We postpone a more detailed discussion about this issue to future work.}
\label{fig:nt6_rescaled_trace}
\end{figure*}

Figure~\ref{fig:nt6_rescaled_trace} shows our results for the temperature dependence of the trace of the energy-momentum tensor $\Delta$, normalized in units of $T^3$ and per gluon. The data shown in this plot are the same that we used to evaluate the pressure in fig.~\ref{fig:nt6_pressure}, hence in the deconfined phase they show the same, approximately perfect, proportionality to $N^2-1$. 
For this observable, however, one also clearly sees that, in the confining phase (in which the number of physical states is independent of $N$---except for the special case $N=2$: see ref.~\cite{Caselle:2011fy} for a discussion), the results corresponding to different gauge groups \emph{do not} follow this proportionality law. This effect was not clearly visible in fig.~\ref{fig:nt6_pressure}, because the pressure and the interaction measure are related to each other by eq.~(\ref{delta_and_p}), and the signal for $p$ at $T<T_c$ is much smaller than at $T>T_c$.

Since $p$ is obtained by numerical integration according to eq.~(\ref{p_lattice}), whereas $\Delta$ is directly related to the plaquette expectation values, see eq.~(\ref{Delta_lattice}), it is most natural to compare the lattice results and the predictions of our holographic model for the interaction measure. This is shown by the orange line in fig.~\ref{fig:nt6_rescaled_trace}, which corresponds to the prediction for $\alpha=3/2$, as discussed in section~\ref{sec:holographic_model}. As one can see, our holographic model accurately captures the non-trivial temperature dependence of $\Delta$ for all temperatures $T \gtrsim 2.5~T_c$. This good quantitative agreement breaks down in the region of temperatures closer to $T_c$, where the holographic curve falls below the lattice results, and the agreement is only qualitative. While in principle the holographic prediction could be adjusted to fit the lattice data over an even broader temperature range (by including more terms in the dilaton potential), we emphasize that, even for the simple setup discussed here, the model already gives a quantitatively correct description for the high-temperature fall-off of $\Delta/[T^3 (N^2-1)]$ and, as discussed in section~\ref{sec:holographic_model}, it relates it to the nature of the deconfinement phase transition.

\begin{figure*}
\centerline{\includegraphics[width=.75\textwidth]{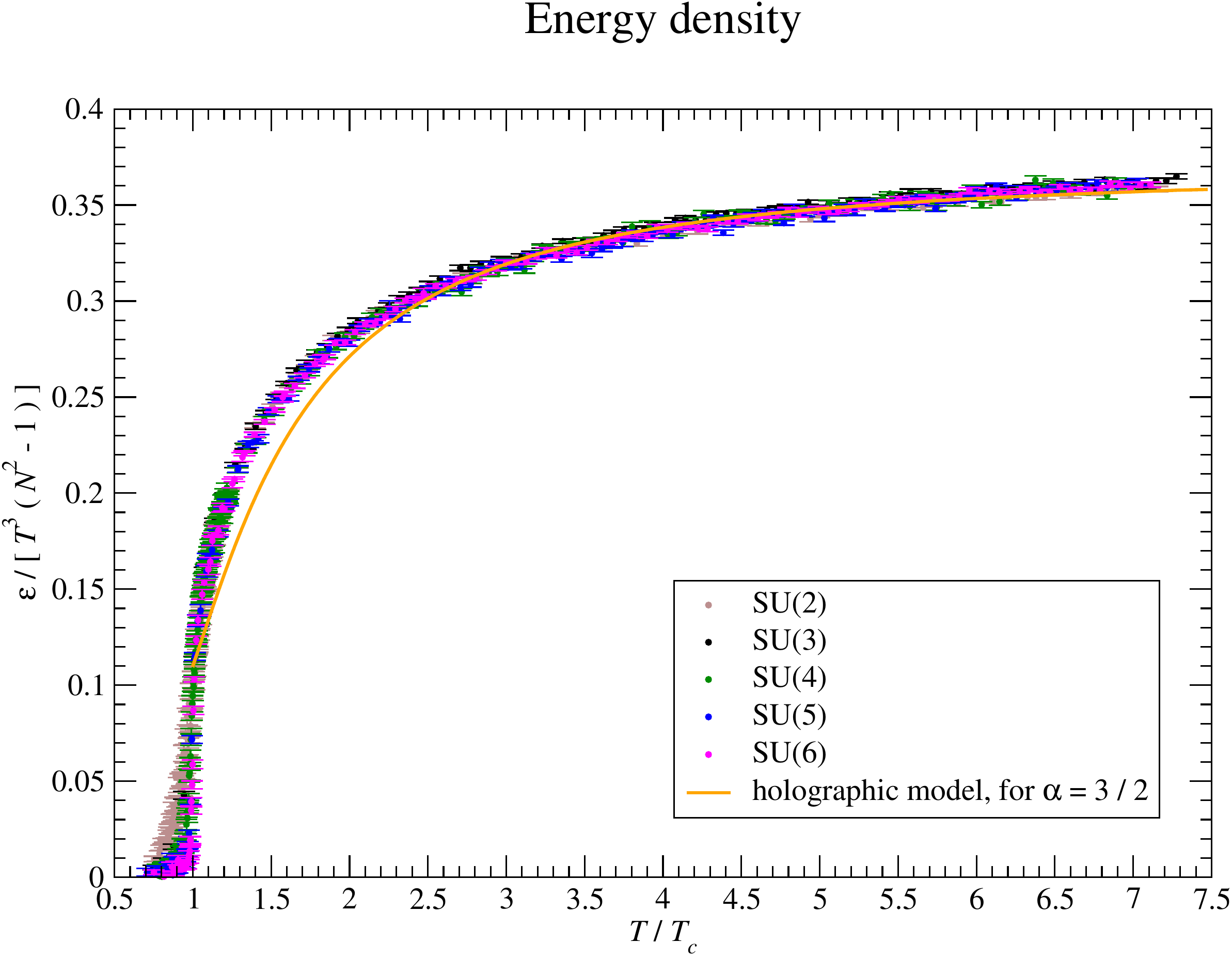}}
\vspace{1cm}
\caption{Same as in fig.~\ref{fig:nt6_pressure}, but for the energy density per gluon, in units of $T^3$.}
\label{fig:nt6_energy}
\end{figure*}

\begin{figure*}
\centerline{\includegraphics[width=.75\textwidth]{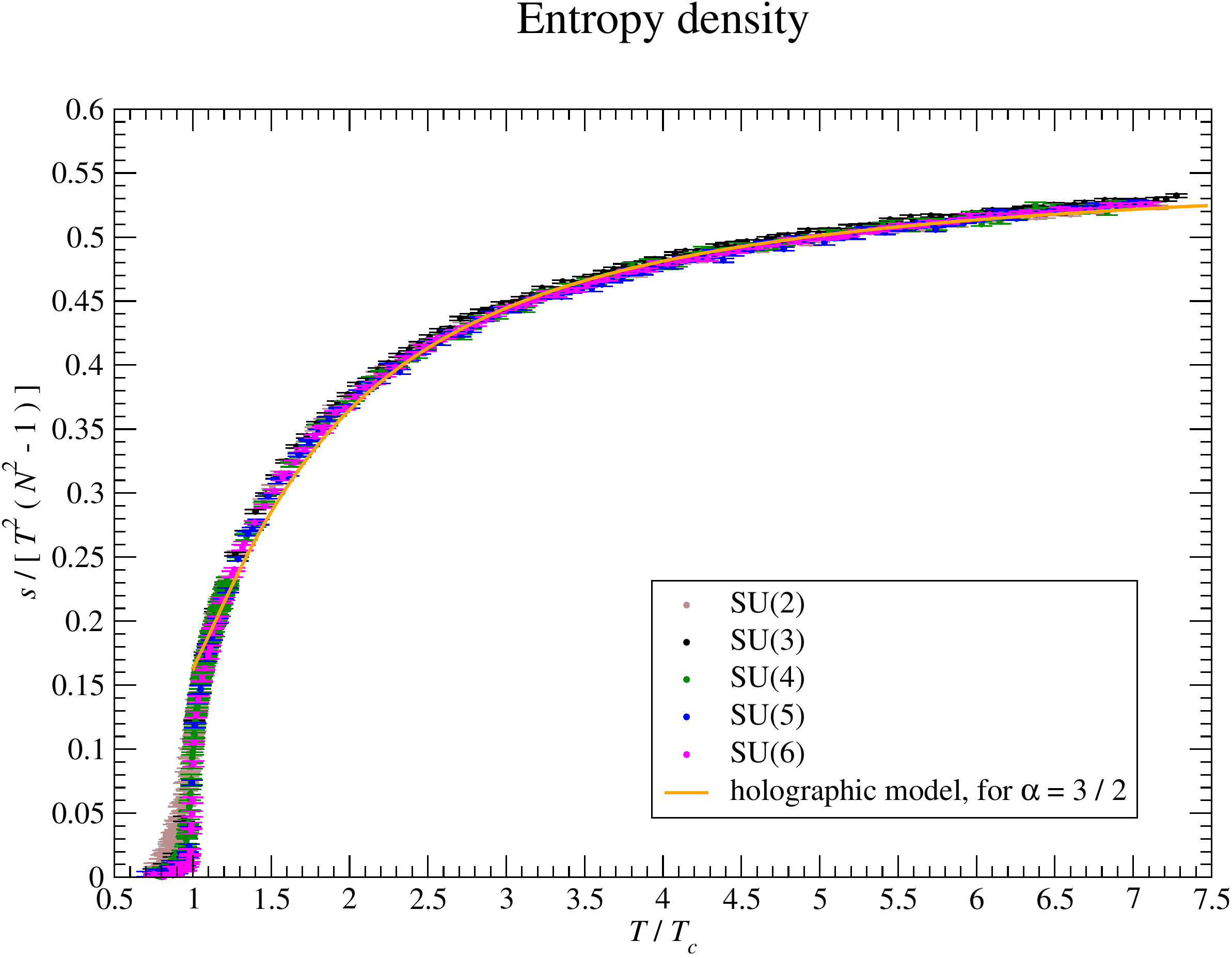}}
\vspace{1cm}
\caption{Same as in fig.~\ref{fig:nt6_pressure}, but for the entropy density per gluon, in units of $T^2$.}
\label{fig:nt6_entropy}
\end{figure*}

Our lattice results for the other two equilibrium thermodynamic quantities (the energy and the entropy density) are shown in fig.~\ref{fig:nt6_energy} and in fig.~\ref{fig:nt6_entropy}, respectively. Being linear combinations of the pressure and the interaction measure, these quantities obviously exhibit the same, accurate proportionality to $N^2-1$ as $p$ and $\Delta$, and reveal the same type of mismatch between lattice data and the holographic prediction for temperatures close to $T_c$. 

\begin{figure*}
\centerline{\includegraphics[width=.75\textwidth]{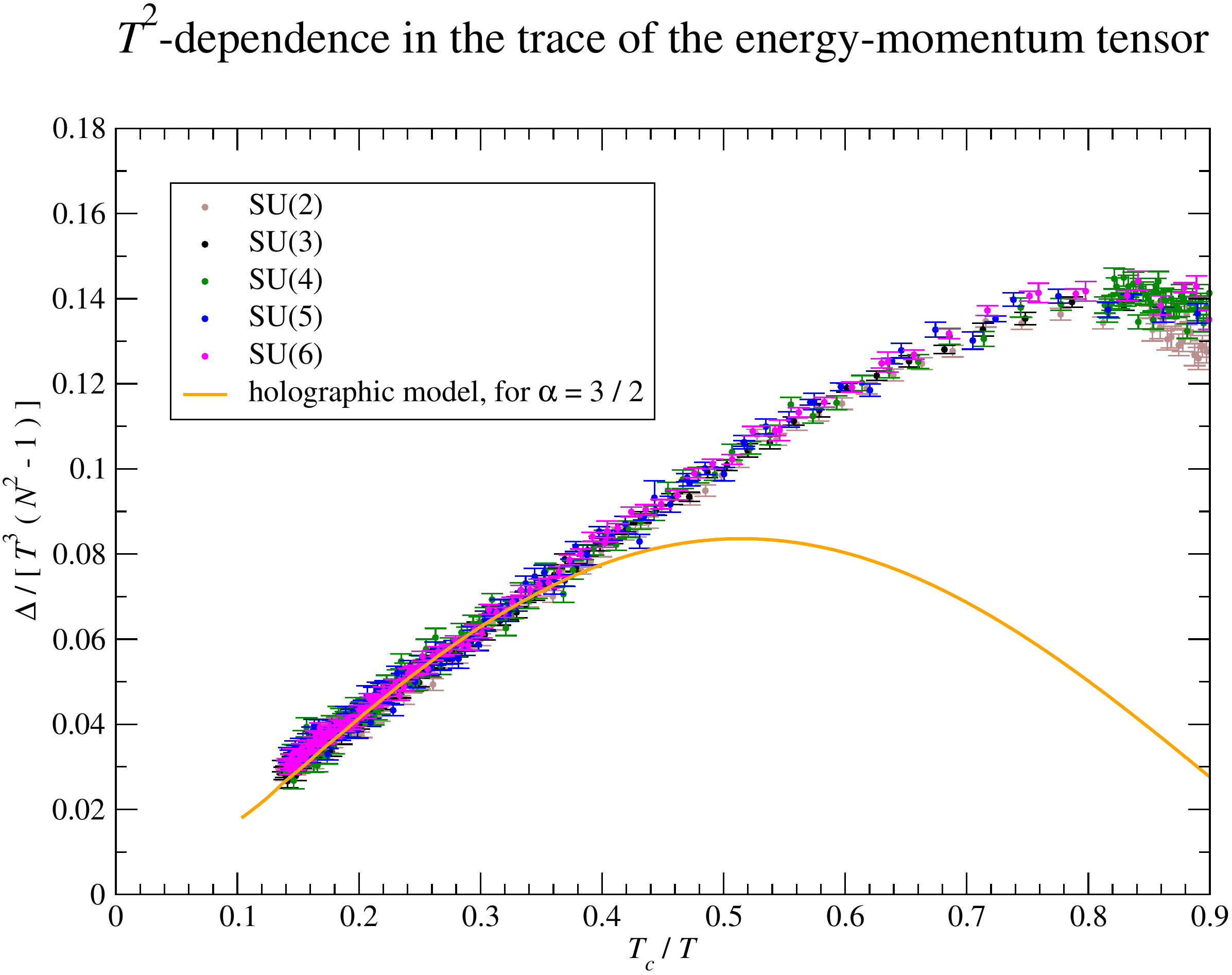}}
\vspace{1cm}
\caption{Similarly to what happens in $D=3+1$ dimensions~\protect\cite{Panero:2009tv}, in the deconfined phase there exists a temperature regime, in which the trace of the energy-momentum tensor $\Delta$ appears to be proportional to $T^2$. This is exhibited very clearly by the linear behavior of the data displayed in this plot, showing the dimensionless ratio $\Delta / [(N^2-1)T^3]$, as a function of $T_c/T$, for temperatures (approximately) starting from $1.1~T_c$ (near the maximum in figure~\ref{fig:nt6_rescaled_trace}). The figure shows the results of our simulations on lattices with $N_t=6$, with the same color code as in fig.~\ref{fig:nt6_pressure}, and the corresponding prediction from the holographic model (solid orange line).}
\label{fig:nt6_Pisarski}
\end{figure*}

Another interesting problem that we investigated in our simulations is the following: In $D=3+1$ dimensions, several authors observed that, in the deconfined phase, the trace of the energy-momentum tensor appears to be proportional to $T^2$ over a rather broad temperature range~\cite{Panero:2009tv,SU3_EoS,Tsquare_in_4D}. Since several different interpretations have been proposed for this phenomenon~\cite{Tsquare_in_4D}, it is interesting to investigate whether a similar effect also occurs in $D=2+1$ dimensions. We address this issue in figure~\ref{fig:nt6_Pisarski}, by showing our results for the dimensionless ratio $\Delta/[T^3(N^2-1)]$, plotted as a function of $T_c/T$: if, in the temperature range under consideration, the interaction measure is dominated by a contribution proportional to $T^2$, this should result in a linear behavior in the plot. This is indeed clearly seen in the figure, hence we confirm that, similarly to the $D=3+1$ case, also in $D=2+1$ dimensions there is a large temperature interval, starting from the value where $\Delta/[T^D(N^2-1)]$ has its maximum, in which the interaction measure of Yang-Mills theories exhibits a quadratic dependence on $T$. This figure also shows that, at least in the temperature range $T \ge 2.5~T_c$, the holographic model captures this type of temperature dependence very well.

The implications of this result are twofold. On the one hand, our finding can be useful to shed light on the nature of the phenomenon in $D=3+1$. In particular, due to the qualitative differences of Yang-Mills theories in $2+1$ \emph{versus} $3+1$ dimensions, our result might help to rule out some mechanisms that have been proposed to explain the phenomenon in $D=3+1$, \emph{if} they are expected to be at work also in the lower-dimensional case. On the other hand, our holographic model leads quite naturally to a power-law decay of $\Delta/[T^D(N^2-1)]$ with the temperature, and, even more interestingly, it suggests a connection between the order of the deconfining phase transition, and the exponent of such power-law decay. As discussed above, the fact that, in general, the deconfinement transition tends towards being more discontinuous when the spacetime dimensionality increases from $2+1$ to $3+1$, can thus be directly related to the change from a $1/T$ to a $1/T^2$ fall-off for $\Delta/[T^D(N^2-1)]$.

\begin{figure*}
\centerline{\includegraphics[width=.45\textwidth]{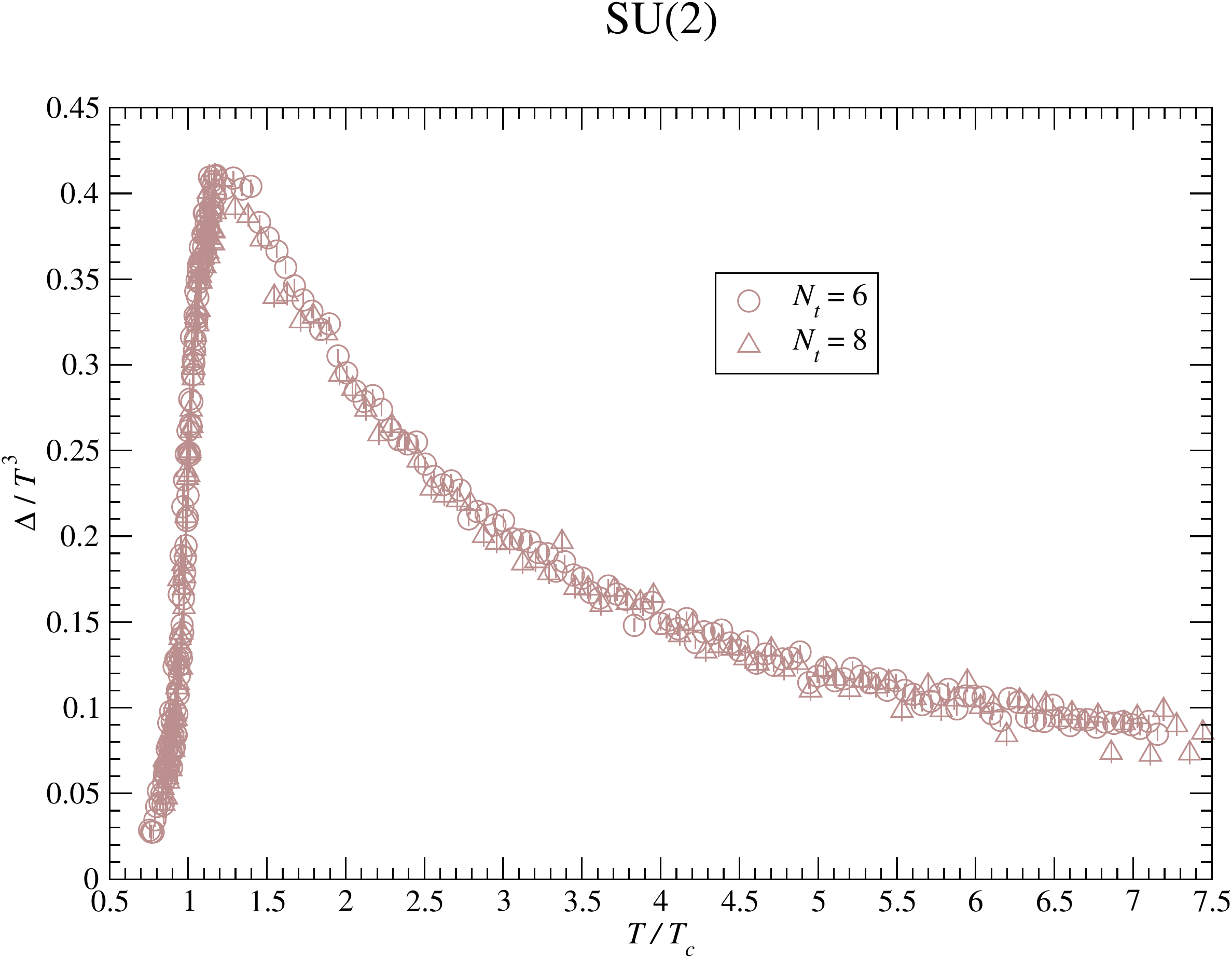} \hfill \includegraphics[width=.45\textwidth]{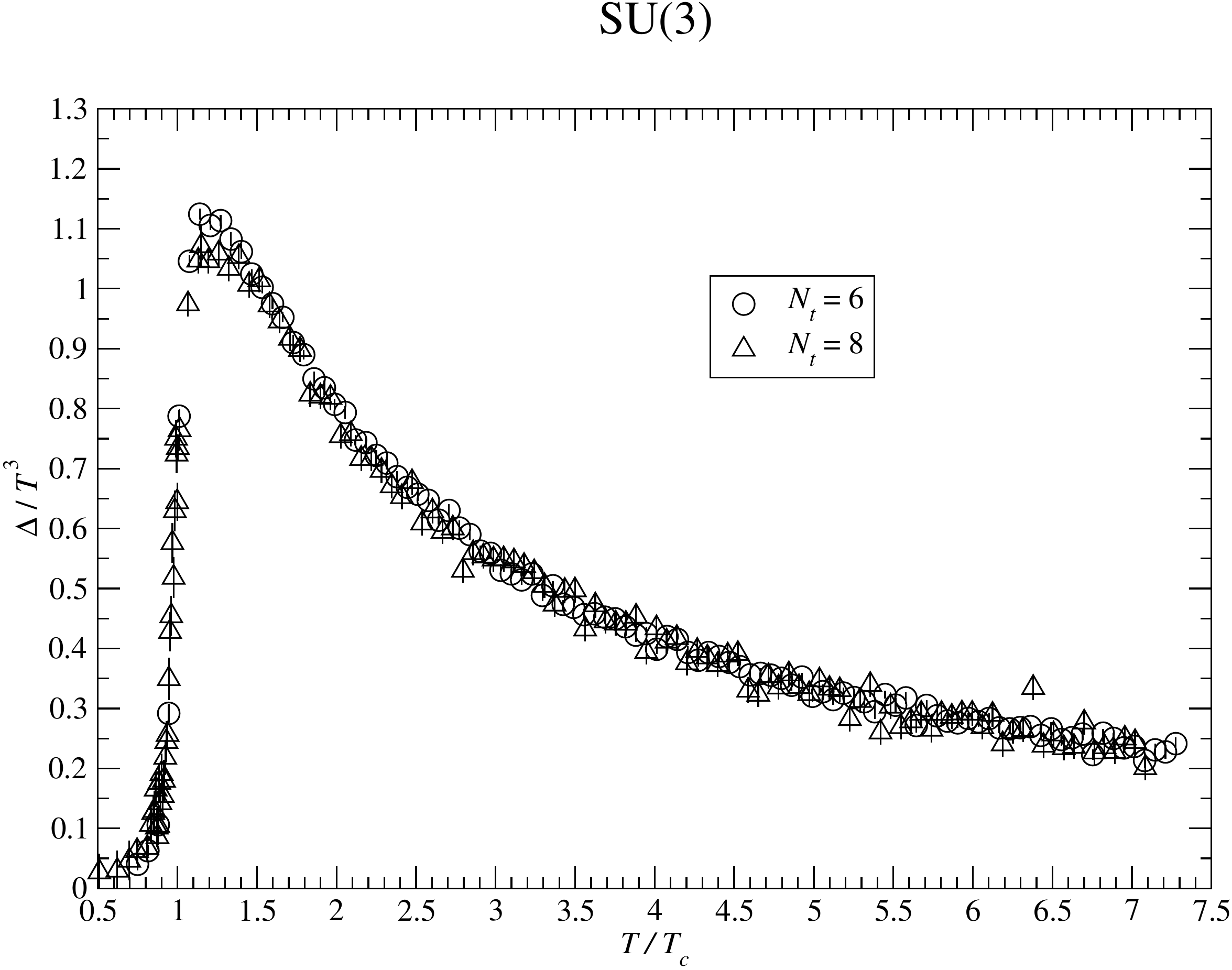}}
\centerline{\includegraphics[width=.45\textwidth]{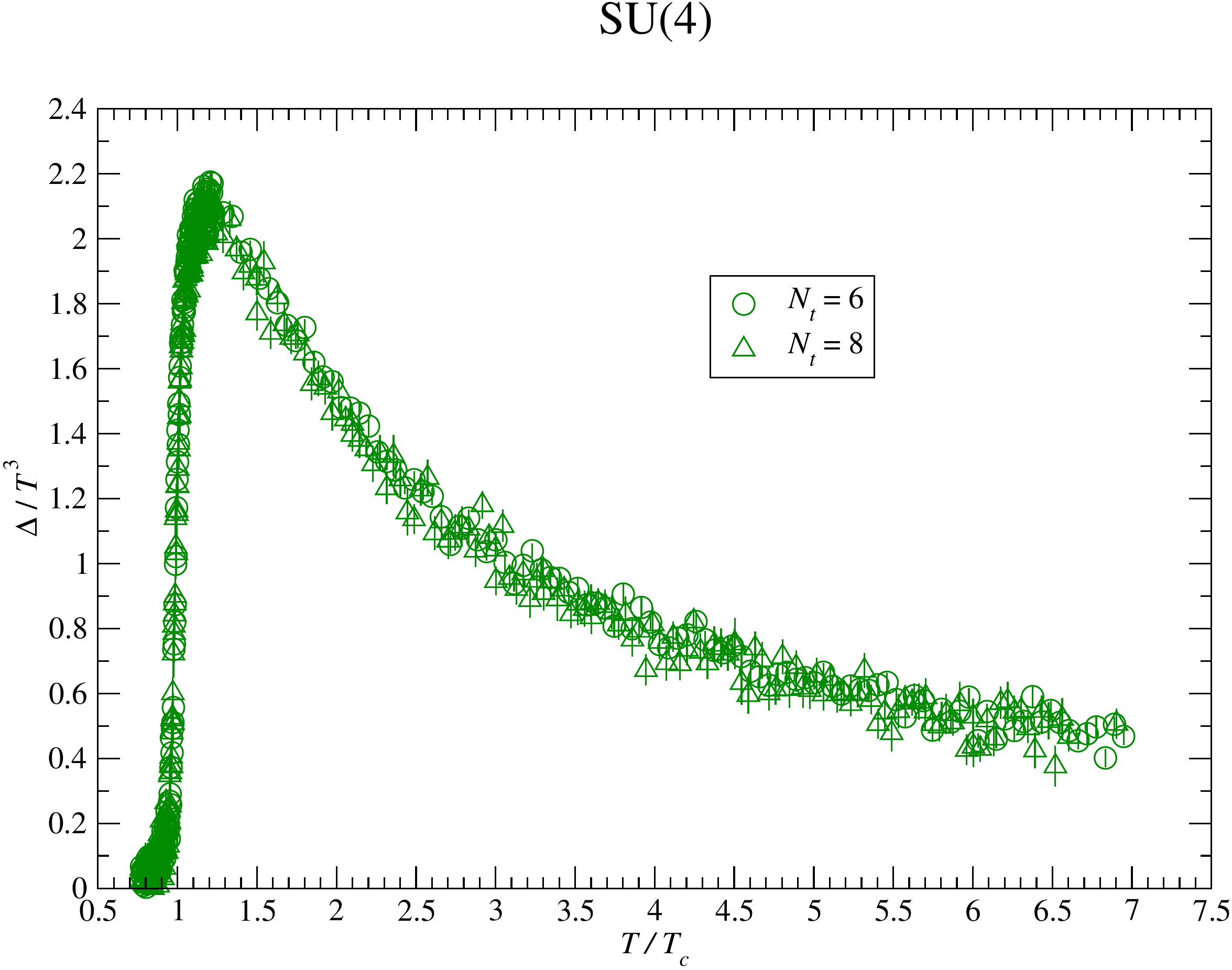} \hfill \includegraphics[width=.45\textwidth]{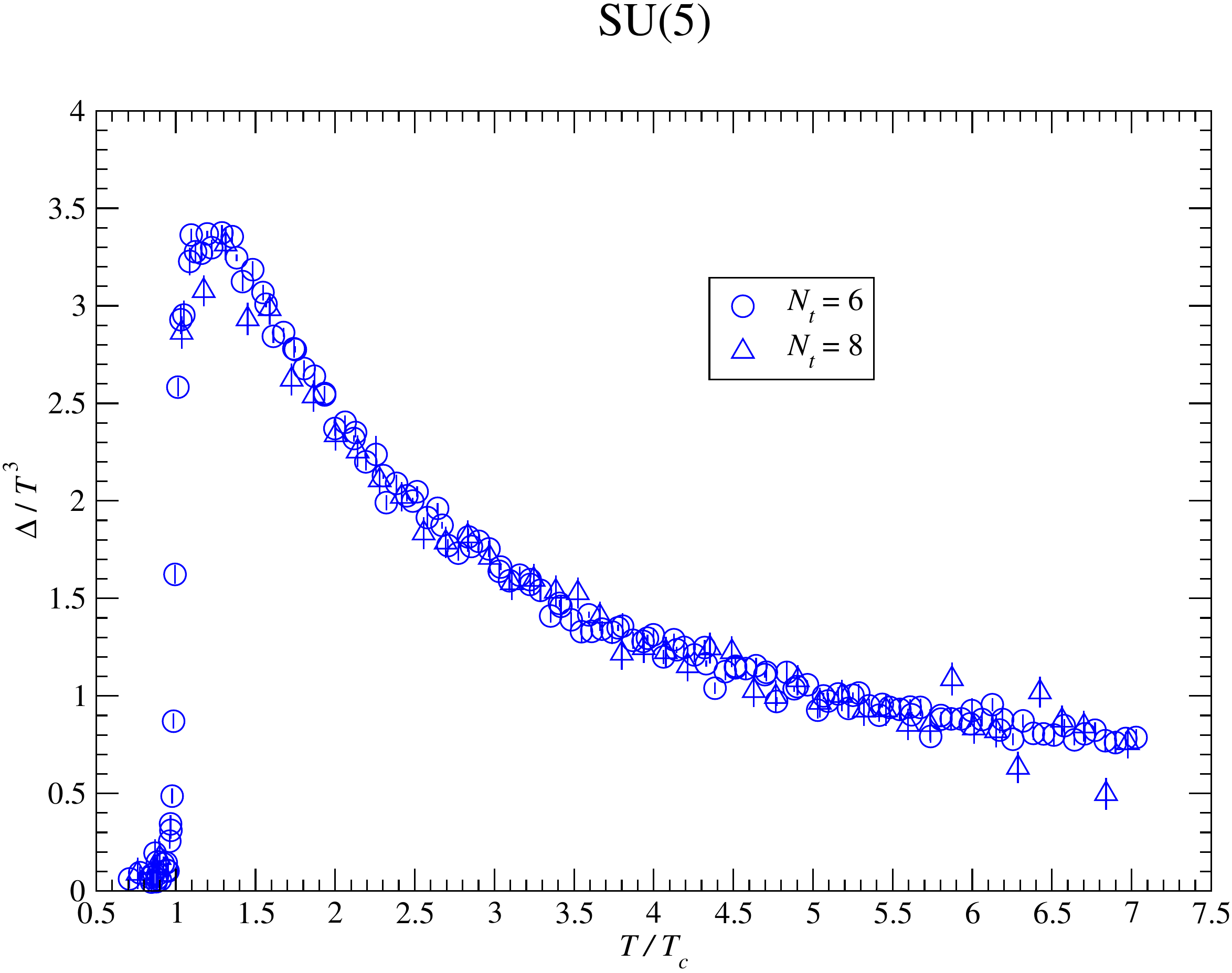}}
\vspace{1cm}
\caption{Cutoff dependence of the trace of the energy-momentum tensor in units of $T^3$: the plots show the results obtained at temperatures ranging from approximately $0.7~T_c$ to $7.5~T_c$, from simulations on lattices with $N_t=6$ (circles) and $8$ (triangles), for four of the gauge groups considered in this work: $\SU(2)$ (top left panel), $\SU(3)$ (top right panel), $\SU(4)$ (bottom left panel) and $\SU(5)$ (bottom right panel).}
\label{fig:Nt_dependence}
\end{figure*}

Finally, we conclude this section with a discussion of the finite-cutoff effects affecting our lattice results. As we mentioned, most of the results presented in this paper are based on finite-temperature simulations using lattices with $N_t=6$ sites in the compactified Euclidean time direction. One may wonder, whether the corresponding results are close enough to the continuum limit or not. According to our analytical expansion of the $\tilde{R}_I(N_t)$ factor in eq.~(\ref{RI_tilde}), it turns out that, for this value of $N_t$, the lattice Stefan-Boltzmann limit (evaluated with the integral method and the Wilson action on an isotropic cubic lattice) differs from the value in the continuum by approximately $5\%$: an effect much larger than the statistical uncertainties and the other systematic errors affecting our data. Thus, in principle one may be tempted to rescale all our lattice results by dividing by $\tilde{R}_I(N_t)$. Since $\tilde{R}_I$ does not depend on $N$, this would not change the fact that the thermodynamic quantities are nearly perfectly proportional to the number of gluons, but would lead to slightly different (smaller) numerical values for $p$, $\Delta$, $\epsilon$ and $s$. However, in the temperature region investigated in this study, this na\"{\i}ve rescaling of the results would not be correct: the physical reason is that the distortion of the Stefan-Boltzmann limit encoded by $\tilde{R}_I$ is due to modes near the lattice cutoff, and those are not relevant for the physics at temperatures of the order of $T_c$. For this reason, we chose not to rescale our numerical results by $\tilde{R}_I$, but rather to repeat our simulations (except for the computationally most demanding gauge group $\SU(6)$) at the same temperatures and at the same space-like volumes, on finer lattices, with $N_t=8$. Given that the leading discretization effects of the Wilson action are $\mathcal{O}(a^2)$, this corresponds to reducing the lattice artifacts by approximately a factor $2$. From eq.~(\ref{p_lattice}) (in which the plaquette mean values are always $\mathcal{O}(1)$, for any $N_t$) it is also easy to see that, when $N_t$ is increased, the difference appearing in the integrand on the right-hand side is affected by a fast decay of the signal-to-noise ratio. As a consequence, it would become extremely difficult to get sufficiently precise results from much finer lattices. Fortunately, however, the discrepancy between our $N_t=6$ and $N_t=8$ results for the equilibrium thermodynamic quantities considered in this work appears to be very small, as figure~\ref{fig:Nt_dependence} shows. This holds in the whole range of temperatures that we studied, and for all values of $N$ from $2$ to $5$. Since our $N_t=6$ and $N_t=8$ data sets yield compatible results, we did not attempt a continuum extrapolation of the thermodynamic quantities, and we can safely state that, to the level of precision we reached, our results from the $N_t=6$ lattices are already compatible with the continuum limit.

\section{Conclusions}
\label{sec:conclusions}

In this work, we presented a non-perturbative study of the equilibrium thermodynamic properties in the deconfined phase of (non-supersymmetric) $\SU(N)$ Yang-Mills theories in $2+1$ dimensions, using holographic computations and numerical simulations based on the lattice regularization. This allowed us to combine the advantages of both tools: the former enables one to gain analytical insight on the dynamical properties of these strongly coupled systems, while the latter (once the thermodynamic and continuum limits are taken) provides numerical results obtained from an \emph{ab initio} approach, directly based on the microscopic definition of the theories for any number of colors, without any assumption or uncontrolled systematic uncertainty.

First, we introduced a holographic bottom-up model, inspired by the IHQCD model~\cite{GK,GKN,Kiritsis}, which describes the non-trivial dynamics of these strongly interacting non-Abelian gauge theories in the large-$N$ limit. This model reveals a non-trivial relationship between the order of the deconfinement phase transition, and the dependence of the trace of the energy-momentum tensor $\Delta$ on the temperature. In particular, for non-Abelian gauge theories in $2+1$ dimensions (which, typically, are characterized by a tendency towards a second-order or a weaker first-order transition than in $3+1$ dimensions), at temperatures of the order of $T_c$ the model favors a behavior approximately compatible with a $1/T$ decay for the dimensionless ratio $\Delta/T^3$.

Then, we defined the non-perturbative regularization of $\SU(N)$ Yang-Mills theories on a $(2+1)$-dimensional Euclidean lattice, and performed a set of high-precision numerical simulations to study their equation of state at $T \ge T_c$. We compared the results obtained for different numbers of colors, up to $N=6$, and found that the trace of the energy-momentum tensor and the related bulk thermodynamic quantities \emph{per gluon} are independent of $N$, reflecting a strikingly accurate scaling of the equation of state in the deconfined phase, over the whole temperature range that we probed (up to about $7.5~T_c$). This holds for all the gauge groups that we studied, including $\SU(2)$, and---at least for these equilibrium thermodynamic observables---supports the potential quantitative relevance of analytical computations relying on the large-$N$ limit (including, in particular, those based on the gauge/gravity correspondence). We also found that, in all the theories that we simulated, $\Delta$ exhibits a clear, characteristic quadratic dependence on the temperature. Both these findings are analogous to those which have been obtained for non-Abelian gauge theories in $3+1$ dimensions~\cite{Panero:2009tv,SU3_EoS,Tsquare_in_4D}. Finally, we compared the lattice results with the prediction of our holographic model, finding good quantitative agreement, at least for temperatures not too close to $T_c$.

In the future, we plan to extend the present study, by investigating the holographic model in more detail, and by looking at different observables, which could be compared with the results of lattice simulations.

\vskip1.0cm {\bf Acknowledgements.}\\
We warmly thank J.~Engels, J.~Erdmenger, A.~O'Bannon, R.~D.~Pisarski and U.~A.~Wiedemann for helpful discussions, comments and correspondence. The numerical simulations were partially performed on the INFN Milano-Bicocca TURING cluster. L.C. acknowledges partial support from DFG (SFB/TR 55) and the European Union grant 238353 (ITN STRONGnet). M.P. acknowledges financial support from the Academy of Finland, project 1134018. This research was supported in part by the National Science Foundation under Grant No. NSF PHY05-51164.

\appendix
\section{Lattice cutoff corrections to the Stefan-Boltzmann limit in $2+1$ and $3+1$ dimensions}
\label{app:lattice_SB}
\renewcommand{\theequation}{A.\arabic{equation}}
\setcounter{equation}{0}

In this appendix, following the calculation in ref.~\cite{Engels:1999tk}, we derive the correction to the Stefan-Boltzmann limit due to cutoff effects on the lattice, for the Wilson discretization of $\SU(N)$ Yang-Mills theory in $D=d+1$ dimensions, for $d=2$ and $3$. Our goal is to evaluate the first few terms of the correction to the Stefan-Boltzmann limit, in an expansion in powers of $N_t^{-2}$, where $N_t$ denotes the number of lattice points in the Euclidean time direction. We take $N_s$, the number of lattice sites along the space-like directions, to be infinite (corresponding to the thermodynamic limit). 

Throughout this appendix, we work in lattice units, i.e., we set the lattice spacing $a$ to unity, and denote the pressure as $p$, the spatial volume as $V$, and the temperature as $T$. Moreover, in the following, we use the $\cong$ notation to mean equality of two quantities, up to terms which are negligible to the order of precision of our computation.

Notation used throughout this calculation includes:
\ba
\omega = \sqrt{ \sum_{i=1}^{d} \sin^2(p_i/2) }, \qquad
x = 2 \arsinh (\omega), \qquad y_i = N_t \sin(p_i/2), \nonumber \\
y = N_t \omega, \qquad t = 2y, \qquad g = -\frac{y^3}{3N_t^2} + \frac{3y^5}{20N_t^4} - \frac{5 y^7}{56 N_t^6}, \qquad 
h = \frac{1}{e^{2y}-1}, \nonumber
\ea
so that:
\be
g^2 \cong \frac{y^6}{9N_t^4} - \frac{y^8}{10 N_t^6}, \qquad
g^3 \cong - \frac{y^9}{27N_t^6}, \qquad \frac{\partial g}{\partial y} = -\frac{y^2}{N_t^2} + \frac{3y^4}{4N_t^4} - \frac{5y^6}{8N_t^6}. \nonumber
\ee

Elementary identities used in this calculation include:

\be
\arcsin (x) = \sum_{k=0}^{\infty} \frac{(2k)!}{2^{2k} (k!)^2} \frac{x^{2k+1}}{2k+1} \qquad \mbox{and} \qquad \arsinh(x) = \sum_{k=0}^{\infty} (-1)^k \frac{(2k)!}{2^{2k} (k!)^2} \frac{x^{2k+1}}{2k+1}, \nonumber
\ee
as well as:
\ba
\Gamma(s+1) \zeta(s+1) &=& \int_0^{\infty} dt \frac{t^s}{e^t-1} \nonumber \\
\frac{1}{2^{s+1}} \Gamma(s+1) \zeta(s+1) &=& \int_0^{\infty} dy \frac{y^s}{e^{2y}-1} \nonumber \\
\Gamma(s+1) \zeta(s) &=& \int_0^{\infty} \!\!\!  dt \; t^s (h+h^2) \nonumber \\
\Gamma(s+1) \zeta(s-1) &=& \int_0^{\infty} \!\!\!  dt  \; t^s (2h^3+3h^2+h) \nonumber \\
\Gamma(s+1) \zeta(s-2) &=& \int_0^{\infty} \!\!\!  dt  \; t^s (6h^4+12h^3+7h^2+h) \nonumber 
\ea
and the expansion:
\ba
\frac{1}{e^{2y+g}-1} &\cong& h + \left( \frac{y^3}{3N_t^2} - \frac{3y^5}{20 N_t^4} + \frac{5y^7}{56 N_t^6}  \right) (h+h^2) + \left( \frac{y^6}{18N_t^4} - \frac{y^8}{20 N_t^6}  \right)(2h^3+3h^2+h)  \nonumber \\
& & \phantom{R}+ \frac{y^9}{162 N_t^6} (6h^4+12h^3+7h^2+h). \nonumber
\ea
Furthermore, we also use the following finite-sum formula~\cite{Elze:1988zs}:
\eq
\sum_{l=1}^{N_t-1} \ln \left[ \omega^2 + \sin^2 \left( \frac{\pi l}{N_t} \right) \right] = 2 \ln \frac{\sinh(N_t x/2)}{2^{N_t - 1} \sinh(x/2) } \nonumber
\en
and the limit:
\eq
\lim_{N_t \to \infty} \frac{1}{N_t} \sum_{l=0}^{N_t-1} \ln \left[ \omega^2 + \sin^2 \left( \frac{\pi l}{N_t} \right) \right] = x - 2 \ln 2, \nonumber
\en
implying:
\be
\frac{1}{N_t} \sum_{l=0}^{N_t-1} \ln \left[ \omega^2 + \sin^2 \left( \frac{\pi l}{N_t} \right) \right] - \lim_{N_t \to \infty} \frac{1}{N_t} \sum_{l=0}^{N_t-1} \ln \left[ \omega^2 + \sin^2 \left( \frac{\pi l}{N_t} \right) \right] = \frac{2}{N_t} \ln \left( 1 - e^{-N_t x} \right). \nonumber
\ee
Finally, in the following we use $ Z^{1 DOF}$ to denote the partition function for one bosonic, massless degree of freedom on the lattice.

We concentrate on the \emph{integral method} for the lattice determination of the pressure~\cite{Engels:1990vr}, in which $p(T)$, the pressure at a given temperature $T$, is defined with respect to its value at $T=0$. The thermodynamic definition of the pressure reads:
\eq
p = T \frac{\partial}{\partial V} \ln Z, \nonumber
\en
and, for an isotropic system, in the thermodynamic limit it reduces to:
\eq
p = \frac{T}{ V} \ln Z. \nonumber
\en

\subsection{$D=3+1$}

The pressure can be written as:
\ba
p &=& 2(N^2-1)\frac{T}{ V} \ln Z^{1 DOF} = \frac{2(N^2-1)}{N_t N_s^3} \ln \left[ \prod_{\vec{p}} \prod_{p_4} \left(
\sum_{\mu=1}^4 \sin^2(p_\mu/2) \right) \right]^{-1/2} \nonumber \\
&=&  -\frac{N^2-1}{N_t} \frac{1}{(2\pi)^3} \int_{([-\pi,\pi])^3} \!\!\! d^3p \sum_{l=0}^{N_t-1} \ln \left[ \omega^2 + \sin^2 \left( \frac{\pi l}{N_t} \right) \right]. \label{eq:D_3+1_pressure}
\ea
Thus, taking into account that in the integral method the pressure is defined w.r.t. to its value at $T=0$ (obtained as the $N_t \to \infty$ limit):
\ba
\frac{p}{T^4} = -2 \frac{(N^2-1)N_t^3}{\pi^3} \int_{([0,\pi])^3} \!\!\! d^3p \ln \left( 1 - e^{-N_t x}\right). \nonumber
\ea
Changing variables to $y_i=N_t \sin (p_i/2)$, and expanding $p_i$ as:
\eq
p_i = 2 \arcsin \left( \frac{y_i}{N_t} \right) \cong 2\frac{y_i}{N_t} + \frac{1}{3} \left( \frac{y_i}{N_t} \right)^3 + \frac{3}{20} \left( \frac{y_i}{N_t} \right)^5 + \frac{5}{56} \left( \frac{y_i}{N_t} \right)^7 , \nonumber
\en
so that:
\eq
\frac{\partial p_i}{\partial y_j} \cong \frac{2}{N_t} \delta_{ij} \left(
1 + \frac{y_i^2}{2N_t^2} + \frac{3y_i^4}{8N_t^4}+ \frac{5y_i^6}{16N_t^6}
\right), \nonumber
\en
one gets:
\ba
\left| \det \left( \frac{\partial p_i}{\partial y_j} \right) \right| \cong \frac{8}{N_t^3} \left\{ 1 + \frac{y^2}{2N_t^2} +\frac{1}{N_t^4} \left[ \frac{3(y_1^4 + y_2^4 + y_3^4 )}{8} +  \frac{y_1^2 y_2^2 + y_1^2 y_3^2 +y_2^2 y_3^2}{4}\right] \right.  \nonumber \\
+\left. \frac{1}{N_t^6} \left[ \frac{5(y_1^6 + y_2^6 + y_3^6 )}{16} + \frac{3(y_1^4y_2^2 + y_1^4 y_3^2 + y_2^4 y_1^2 + y_2^4 y_3^2 + y_3^4 y_1^2 + y_3^4 y_2^2 )}{16} +\frac{y_1^2 y_2^2 y_3^2 }{8} \right] \right\} \nonumber
\ea
and therefore (by rotational symmetry):
\ba
d^3p \cong d^3y \cdot \frac{8}{N_t^3} \left( 1 +  \frac{y^2}{2N_t^2} +  \frac{9 y_3^4 + 6 y_1^2 y_3^2}{8N_t^4} +  \frac{15 y_3^6 + 18 y_1^2 y_3^4 + 2 y_1^2 y_2^2 y_3^2}{16N_t^6} 
\right) . \label{eq:d3p_to_d3y}
\ea
Thus the dimensionless ratio $p/T^4$ can be written as:
\ba
\frac{p}{T^4} &\cong& -2 \frac{N^2-1}{\pi^3} \int_{([-N_t,N_t])^3} \!\!\! d^3y  \left[ 1 + \frac{y^2}{2N_t^2} +  \frac{1}{8N_t^4} (9 y_3^4 + 6 y_1^2 y_3^2) \right.  \nonumber \\
&&  \left. +   \frac{1}{16N_t^6} (15 y_3^6 + 18 y_1^2 y_3^4 + 2 y_1^2 y_2^2 y_3^2)  \right] \cdot \ln \left( 1 - e^{-N_t x} \right)       \nonumber \\
&\cong& 
-2 \frac{N^2-1}{\pi^3}\int_{\R^3} \!\!\! d^3y\ln\left(1-e^{-N_t x}\right)     
  -\frac{N^2-1}{N_t^2\pi^3}\int_{\R^3} \!\!\! d^3y  \; y^2 \ln\left(1-e^{-N_t x}\right) \nonumber \\
&&-\frac{9(N^2-1)}{4N_t^4 \pi^3}\int_{\R^3} \!\!\! d^3y \;  y_3^4\ln\left(1-e^{-N_t x}\right)  
-\frac{3(N^2-1)}{2N_t^4 \pi^3}\int_{\R^3} \!\!\! d^3y  \; y_1^2 y_3^2\ln\left(1-e^{-N_t x}\right) \nonumber \\
&&-\frac{15(N^2-1)}{8N_t^6 \pi^3}\int_{\R^3} \!\!\! d^3y  \; y_3^6 \ln\left(1-e^{-N_t x}\right) 
-\frac{9(N^2-1)}{4N_t^6 \pi^3}\int_{\R^3} \!\!\! d^3y  \; y_1^2 y_3^4 \ln\left(1-e^{-N_t x}\right) \nonumber \\
&& -\frac{N^2-1}{4 N_t^6 \pi^3}\int_{\R^3} \!\!\! d^3y  \; y_1^2 y_2^2 y_3^2\ln\left(1-e^{-N_t x}\right).
\label{eq:D4expansion}
\ea
The last expression can be readily evaluated, using the formulas listed above and integration by parts. In particular, the first two terms are:
\be
-2 \frac{N^2-1}{\pi^3}\int_{\R^3} \!\!\! d^3y\ln\left(1-e^{-N_t x}\right) \cong \frac{N^2-1}{\pi^2} \left[ 2 \zeta(4) + \frac{5}{N_t^2} \zeta(6) + \frac{91}{8N_t^4} \zeta(8) + \frac{205}{8N_t^6} \zeta(10)\right] \nonumber
\ee 
and:
\be
-\frac{N^2-1}{N_t^2\pi^3}\int_{\R^3} \!\!\! d^3y  \; y^2 \ln\left(1-e^{-N_t x}\right) \cong \frac{N^2-1}{\pi^2} \left[ \frac{3}{N_t^2} \zeta(6) + \frac{105}{4N_t^4} \zeta(8) + \frac{1449}{8N_t^6} \zeta(10)\right]. \nonumber
\ee
Next, note that, introducing the following polar parametrization for the $y_i$ coordinates:
\be
\left\{
\begin{array}{l}
y_1 = y \sin \theta \cos \phi \\
y_2 = y \sin \theta \sin \phi \\
y_3 = y \cos \theta
\end{array}
\right.
\nonumber
\ee
and denoting $c=\cos\theta$, one gets:
\ba
-\frac{9(N^2-1)}{4N_t^4 \pi^3}\int_{\R^3} \!\!\! d^3y \;  y_3^4\ln\left(1-e^{-N_t x}\right)   \cong - \frac{9(N^2-1)}{2\pi^2N_t^4} \int_{-1}^{1} \!\!\! dc  \; c^4 \int_0^{\infty}  \!\!\! dy  \; y^6 \ln \left( 1 -e^{-2y-g} \right) \nonumber \\
 \cong \frac{18(N^2-1)}{35\pi^2N_t^4}   \int_0^{\infty}  \!\!\! dy \frac{y^7}{e^{2y+g}-1} \left[ 1  -\frac{y^2}{2N_t^2} \right] \cong  81\frac{N^2-1}{8\pi^2} \left[ \frac{1}{N_t^4} \zeta(8) + \frac{21}{N_t^6} \zeta(10)\right]. \nonumber
\ea
Similarly:
\ba
&& -\frac{3(N^2-1)}{2N_t^4 \pi^3}\int_{\R^3} \!\!\! d^3y  \; y_1^2 y_3^2\ln\left(1-e^{-N_t x}\right) \nonumber \\
&\cong& - \frac{3(N^2-1)}{2\pi^2N_t^4} \int_0^{2\pi}  \!\!\! d \phi  \; \cos^2\phi \int_{-1}^{1} \!\!\! dc  \; c^2 (1-c^2) \int_0^{\infty}  \!\!\! dy  \; y^6 \ln \left( 1 -e^{-2y-g} \right) \nonumber \\
&\cong& 9\frac{N^2-1}{4\pi^2} \left[ \frac{1}{N_t^4} \zeta(8) + \frac{21}{N_t^6} \zeta(10)\right]. \nonumber
\ea
The remaining terms evaluate to:
\ba
&& -\frac{15(N^2-1)}{8N_t^6 \pi^3}\int_{\R^3} \!\!\! d^3y  \; y_3^6 \ln\left(1-e^{-N_t x}\right)  \nonumber \\
&\cong& -\frac{15(N^2-1)}{4 \pi^2N_t^6} \int_{-1}^{1}  \!\!\! dc  \; c^6 \int_0^{\infty}  \!\!\! dy  \; y^8 \ln \left( 1 -e^{-2y-g} \right) \nonumber \\
&\cong& \frac{N^2-1}{8\pi^2} \left[ \frac{675}{N_t^6} \zeta(10) \right], \nonumber
\ea
to:
\ba
&&-\frac{9(N^2-1)}{4N_t^6 \pi^3}\int_{\R^3} \!\!\! d^3y  \; y_1^2 y_3^4 \ln\left(1-e^{-N_t x}\right) \nonumber \\
&\cong& -\frac{9(N^2-1)}{4 \pi^3N_t^6} \int_0^{2\pi}  \!\!\! d\phi  \; \cos^2 \phi \int_{-1}^{1}  \!\!\! dc  \; c^4 (1-c^2) \int_0^{\infty}  \!\!\! dy  \; y^8 \ln \left( 1 -e^{-2y-g} \right) \nonumber \\
&\cong& \frac{N^2-1}{4\pi^2} \left[ \frac{81}{N_t^6} \zeta(10) \right]\nonumber
\ea
and finally:
\ba
&& -\frac{N^2-1}{4 N_t^6 \pi^3}\int_{\R^3} \!\!\! d^3y  \; y_1^2 y_2^2 y_3^2\ln\left(1-e^{-N_t x}\right) \nonumber \\ &\cong&  -\frac{N^2-1}{4 \pi^3N_t^6} \int_0^{2\pi}  \!\!\! d \phi  \; \cos^2 \phi \sin^2 \phi \int_{-1}^{1}  \!\!\! dc  \; c^2 (1-c^2)^2 \int_0^{\infty}  \!\!\! dy  \; y^8 \ln \left( 1 -e^{-2y-g} \right) \nonumber \\
&\cong& \frac{N^2-1}{4\pi^2} \left[ \frac{3}{N_t^6} \zeta(10) \right].\nonumber
\ea
Plugging these results into eq.~(\ref{eq:D4expansion}), one eventually ends up with:
\be
\frac{p}{T^4} \cong  \frac{N^2-1}{\pi^2} \left[ 2 \zeta(4) + \frac{8}{N_t^2} \zeta(6) + \frac{50}{N_t^4} \zeta(8) + \frac{572}{N_t^6} \zeta(10)\right].  \label{eq:D4expansion_intermediateform}
\ee
Noting that:
\be
\zeta(4)= \frac{\pi^4}{90}, \qquad 
\zeta(6)= \frac{\pi^6}{945}, \qquad 
\zeta(8)= \frac{\pi^8}{9450}, \qquad \mbox{and} \qquad
\zeta(10)= \frac{\pi^{10}}{93555}, \nonumber
\ee
eq.~(\ref{eq:D4expansion_intermediateform}) can be rewritten as:
\eq
\frac{p}{T^4} = \frac{\pi^2}{45} (N^2-1) \cdot R_I(N_t),
\en
where:
\eq
R_I(N_t) = 1 + \frac{8}{21} \left( \frac{\pi}{N_t} \right)^2 + \frac{5}{21} \left( \frac{\pi}{N_t} \right)^4 + \frac{52}{189} \left( \frac{\pi}{N_t} \right)^6 + \mathcal{O}\left( (\pi/N_t)^8 \right),
\en
which reproduces the expression given in ref.~\cite{Engels:1999tk}, and extends it to the next order in powers of $( \pi/N_t)^2$.

\subsection{$D=2+1$}

An analogous calculation can be done for a system in $D=2+1$ dimensions, for which eq.~(\ref{eq:D_3+1_pressure}) gets replaced by:
\ba
p &=& (N^2-1)\frac{T}{ V} \ln Z^{1 DOF} \nonumber \\
&=& \frac{N^2-1}{N_t N_s^2} \ln \left[ \prod_{\vec{p}} \prod_{p_3} \left(
\sum_{\mu=1}^3 \sin^2(p_\mu/2) \right) \right]^{-1/2} \nonumber \\
&=&  -\frac{N^2-1}{2N_t} \frac{1}{(2\pi)^2} \int_{([-\pi,\pi])^2} d^2p \sum_{l=0}^{N_t-1} \ln \left[ \omega^2 + \sin^2 \left( \frac{\pi l}{N_t} \right) \right], \label{eq:D=2+1_pressure}
\ea
while eq.~(\ref{eq:d3p_to_d3y}) gets replaced by:
\ba
d^2p \cong d^2y \cdot \frac{4}{N_t^2} \left( 1 +  \frac{y^2}{2N_t^2} +  \frac{3 y_1^4 + y_1^2 y_2^2}{4N_t^4} +  \frac{5 y_1^6 + 3 y_1^4 y_2^2}{8N_t^6}
\right). \label{eq:d2p_to_d2y}
\ea
Accordingly, the Stefan-Boltzmann limit for the dimensionless ratio $p/T^3$, as evaluated on a finite lattice using the integral method, reads:
\ba
&&\frac{p}{T^3} = - \frac{(N^2-1)N_t^2}{\pi^2} \int_{([0,\pi])^2}  \!\!\! d^2p \ln \left( 1 - e^{-N_t x}\right) \nonumber \\
&\cong&  - \frac{N^2-1}{\pi^2} \int_{([-N_t,N_t])^2}  \!\!\! d^2y  \left[ 1 +  \frac{y^2}{2N_t^2} +  \frac{3 y_1^4 + y_1^2 y_2^2}{4N_t^4} +  \frac{5 y_1^6 + 3 y_1^4 y_3^2}{8N_t^6} \right] \ln \left( 1 - e^{-N_t x}\right) \nonumber \\
&\cong&  -\frac{N^2-1}{\pi^2}\int_{\R^2} \!\!\! d^2y\ln\left(1-e^{-N_t x}\right)
   -\frac{N^2-1}{2N_t^2\pi^2}\int_{\R^2} \!\!\! d^2y  \; y^2 \ln\left(1-e^{-N_t x}\right) \nonumber \\
&& -\frac{3(N^2-1)}{4N_t^4 \pi^2}\int_{\R^2} \!\!\! d^2y \;  y_1^4\ln\left(1-e^{-N_t x}\right) -\frac{N^2-1}{4N_t^4 \pi^2}\int_{\R^2} \!\!\! d^2y  \; y_1^2 y_2^2\ln\left(1-e^{-N_t x}\right)\nonumber \\
&& -\frac{5(N^2-1)}{8N_t^6 \pi^2}\int_{\R^2} \!\!\! d^2y  \; y_1^6 \ln\left(1-e^{-N_t x}\right) -\frac{3(N^2-1)}{8N_t^6 \pi^2}\int_{\R^2} \!\!\! d^2y  \; y_1^4 y_2^2 \ln\left(1-e^{-N_t x}\right).
\label{eq:p_over_T3_expansion}
\ea
Similarly to the $D=3+1$ case, each term appearing in the last expression can be evaluated separately. In particular, one easily finds that:
\be
 -\frac{N^2-1}{\pi^2}\int_{\R^2} \!\!\! d^2y\ln\left(1-e^{-N_t x}\right) \cong \frac{N^2-1}{2\pi} \left[ \zeta(3) + \frac{1}{N_t^2} \zeta(5) +\frac{1}{N_t^4} \zeta(7) + \frac{1}{N_t^6} \zeta(9)\right]
\nonumber
\ee
and
\be
 -\frac{N^2-1}{2N_t^2\pi^2}\int_{\R^2} \!\!\! d^2y  \; y^2 \ln\left(1-e^{-N_t x}\right) \cong 3\frac{N^2-1}{8\pi} \left[ \frac{1}{N_t^2} \zeta(5) + \frac{5}{N_t^4} \zeta(7) + \frac{21}{N_t^6} \zeta(9) \right].
\nonumber
\ee
Next, introducing the following polar parametrization for the $y_i$ coordinates:
\be
\left\{
\begin{array}{l}
y_1 = y \cos \theta \\
y_2 = y \sin \theta
\end{array}
\right. , \nonumber
\ee
it is easy to prove that:
\ba
&& -\frac{3(N^2-1)}{4N_t^4 \pi^2}\int_{\R^2} \!\!\! d^2y \;  y_1^4\ln\left(1-e^{-N_t x}\right) \cong - \frac{3(N^2-1)}{4\pi^2N_t^4} \int_{0}^{2\pi} \!\!\! d\theta  \;  \cos^4\theta \int_0^{\infty}  \!\!\! dy  \; y^5 \ln \left( 1 -e^{-2y-g} \right) \nonumber \\
 &\cong&  135 \frac{N^2-1}{128\pi} \left[ \frac{1}{N_t^4} \zeta(7) + \frac{14}{N_t^6} \zeta(9)\right], \nonumber
\ea
while:
\ba
&& -\frac{N^2-1}{4N_t^4 \pi^2}\int_{\R^2} \!\!\! d^2y  \; y_1^2 y_2^2\ln\left(1-e^{-N_t x}\right) \cong - \frac{N^2-1}{4\pi^2N_t^4} \int_{0}^{2\pi} \!\!\! d\theta \;  \cos^2\theta \sin^2\theta \int_0^{\infty}  \!\!\! dy  \; y^5 \ln \left( 1 -e^{-2y-g} \right) \nonumber \\
 &\cong&  15 \frac{N^2-1}{128\pi} \left[ \frac{1}{N_t^4} \zeta(7) + \frac{14}{N_t^6} \zeta(9)\right]. \nonumber
\ea
Similarly:
\ba
&& -\frac{5(N^2-1)}{8N_t^6 \pi^2}\int_{\R^2} \!\!\! d^2y  \; y_1^6 \ln\left(1-e^{-N_t x}\right)\cong - \frac{5(N^2-1)}{8\pi^2N_t^6} \int_{0}^{2\pi} \!\!\! d\theta \;  \cos^6\theta \int_0^{\infty}  \!\!\! dy  \; y^7 \ln \left( 1 -e^{-2y-g} \right) \nonumber \\
&\cong&   \frac{N^2-1}{1024\pi} \left[ \frac{7875}{N_t^6} \zeta(9) \right] \nonumber
\ea
and:
\ba
&& -\frac{3(N^2-1)}{8N_t^6 \pi^2}\int_{\R^2} \!\!\! d^2y  \; y_1^4 y_2^2 \ln\left(1-e^{-N_t x}\right) \nonumber \\
&\cong& - \frac{3(N^2-1)}{8\pi^2N_t^6} \int_{0}^{2\pi} \!\!\! d\theta  \; \cos^4\theta \sin^2\theta \int_0^{\infty}  \!\!\! dy  \; y^7 \ln \left( 1 -e^{-2y-g} \right) \nonumber \\
 &\cong&   \frac{N^2-1}{1024\pi} \left[ \frac{945}{N_t^6} \zeta(9) \right], \nonumber
\ea
so that the final result reads:
\eq
\frac{p}{T^3} \cong \frac{N^2-1}{\pi} \left[ \frac{1}{2} \zeta(3) +  \frac{7}{8}\frac{1}{N_t^2} \zeta(5) + \frac{227}{64}\frac{1}{N_t^4} \zeta(7) + \frac{8549}{256}\frac{1}{N_t^6} \zeta(9) \right].
\label{eq:p_T3_intermediate}
\en
Eq.~(\ref{eq:p_T3_intermediate}) can be recast in the form:
\eq
\frac{p}{T^3} = \frac{N^2-1}{2\pi} \zeta(3) \cdot \tilde{R}_I(N_t),
\en
with:
\eq
\tilde{R}_I(N_t) = 1 + \frac{7}{4} \frac{1}{N_t^2} \frac{\zeta(5)}{\zeta(3)} + \frac{227}{32} \frac{1}{N_t^4} \frac{\zeta(7)}{\zeta(3)} + \frac{8549}{128} \frac{1}{N_t^6} \frac{\zeta(9)}{\zeta(3)} + \mathcal{O}\left( N_t^{-8} \right).
\en
In contrast to the $D=3+1$ case, the latter expression cannot be rewritten as a simple power series in $( \pi/N_t)^2$ with rational coefficients, because, for odd values of $x$, the $\pi^x/\zeta(x)$ ratio is not just an integer (or a rational) number. However, note that, defining:
\eq
S_{\pm}(n) = \sum_{k=1}^{\infty} \frac{1}{k^n \left( e^{2\pi k} \pm1 \right)}, \nonumber
\en
it is possible to write:
\ba
\zeta(3) &=& \frac{7}{180}\pi^3 -2S_{-}(3) \simeq 1.20205690316\dots \nonumber \\
\zeta(5) &=& \frac{1}{294} \pi^5 -\frac{72}{35} S_{-}(5) -\frac{2}{35} S_{+}(5) \simeq 1.0369277551\dots \nonumber \\
\zeta(7) &=& \frac{19}{56700}\pi^7 -2S_{-}(7) \simeq 1.0083492774\dots \nonumber \\
\zeta(9) &=& \frac{125}{3704778} \pi^9 -\frac{992}{495} S_{-}(9) -\frac{2}{495} S_{+}(9) \simeq 1.0020083928\dots \nonumber 
\ea


\begin{thebibliography}{0}

\bibitem{finiteTexperiments}
%\cite{Heinz:2000bk}
% \bibitem{Heinz:2000bk}
  U.~W.~Heinz and M.~Jacob,
  %``Evidence for a new state of matter: An assessment of the results from  the
  %CERN lead beam programme,''
  arXiv:nucl-th/0002042.
  %%CITATION = NUCL-TH/0002042;%%
% 
%\cite{Gyulassy:2004zy}
% \bibitem{Gyulassy:2004zy}
  M.~Gyulassy and L.~McLerran,
  %``New forms of QCD matter discovered at RHIC,''
  Nucl.\ Phys.\  A {\bf 750} (2005) 30
  [arXiv:nucl-th/0405013].
  %%CITATION = NUPHA,A750,30;%%
% 
%\cite{Adcox:2004mh}
% \bibitem{Adcox:2004mh}
  K.~Adcox {\it et al.}  [PHENIX Collaboration],
  %``Formation of dense partonic matter in relativistic nucleus nucleus
  %collisions at RHIC: Experimental evaluation by the PHENIX  collaboration,''
  Nucl.\ Phys.\  A {\bf 757}, 184 (2005)
  [arXiv:nucl-ex/0410003].
  %%CITATION = NUPHA,A757,184;%%
% 
%\cite{Arsene:2004fa}
% \bibitem{Arsene:2004fa}
  I.~Arsene {\it et al.}  [BRAHMS Collaboration],
  %``Quark Gluon Plasma an Color Glass Condensate at RHIC? The perspective from
  %the BRAHMS experiment,''
  Nucl.\ Phys.\  A {\bf 757}, 1 (2005)
  [arXiv:nucl-ex/0410020].
  %%CITATION = NUPHA,A757,1;%%
% 
%\cite{Back:2004je}
% \bibitem{Back:2004je}
  B.~B.~Back {\it et al.},
  %``The PHOBOS perspective on discoveries at RHIC,''
  Nucl.\ Phys.\  A {\bf 757}, 28 (2005)
  [arXiv:nucl-ex/0410022].
  %%CITATION = NUPHA,A757,28;%%
% 
%\cite{Adams:2005dq}
% \bibitem{Adams:2005dq}
  J.~Adams {\it et al.}  [STAR Collaboration],
  %``Experimental and theoretical challenges in the search for the quark  gluon
  %plasma: The STAR collaboration's critical assessment of the  evidence from
  %RHIC collisions,''
  Nucl.\ Phys.\  A {\bf 757}, 102 (2005)
  [arXiv:nucl-ex/0501009].
  %%CITATION = NUPHA,A757,102;%%
% 
%\cite{Aad:2010bu}
% \bibitem{Aad:2010bu}
  G.~Aad {\it et al.}  [Atlas Collaboration],
  %``Observation of a Centrality-Dependent Dijet Asymmetry in Lead-Lead
  %Collisions at sqrt(S(NN))= 2.76 TeV with the ATLAS Detector at the LHC,''
  Phys.\ Rev.\ Lett.\  {\bf 105} (2010) 252303
  [arXiv:1011.6182 [hep-ex]].
  %%CITATION = PRLTA,105,252303;%%
% 
%\cite{Chatrchyan:2011sx}
% \bibitem{Chatrchyan:2011sx}
  S.~Chatrchyan {\it et al.} [ CMS Collaboration ],
  %``Observation and studies of jet quenching in PbPb collisions at nucleon-nucleon center-of-mass energy = 2.76 TeV,''
  Phys.\  Rev.\  {\bf C84 } (2011)  024906
  [arXiv:1102.1957 [nucl-ex]].
  %%CITATION = ARXIV:1102.1957;%%
% 
%\cite{Aamodt:2010pa}
% \bibitem{Aamodt:2010pa}
  K.~Aamodt {\it et al.}  [ALICE Collaboration],
  %``Elliptic flow of charged particles in Pb-Pb collisions at 2.76 TeV,''
  Phys.\ Rev.\ Lett.\  {\bf 105} (2010) 252302
  [arXiv:1011.3914 [nucl-ex]];
  %%CITATION = ARXIV:1011.3914;%%
% 
%\cite{Aamodt:2010pb}
% \bibitem{Aamodt:2010pb}
%   K.~Aamodt {\it et al.}  [ALICE Collaboration],
  %``Charged-particle multiplicity density at mid-rapidity in central Pb-Pb
  %collisions at sqrt(sNN) = 2.76 TeV,''
  Phys.\ Rev.\ Lett.\  {\bf 105} (2010) 252301
  [arXiv:1011.3916 [nucl-ex]];
  %%CITATION = PRLTA,105,252301;%%
% 
%\cite{Aamodt:2010jd}
% \bibitem{Aamodt:2010jd}
%   K.~Aamodt {\it et al.}  [ALICE Collaboration],
  %``Suppression of Charged Particle Production at Large Transverse Momentum in
  %Central Pb--Pb Collisions at $\sqrt{s_{_{NN}}} = 2.76$ TeV,''
  Phys.\ Lett.\  B {\bf 696} (2011) 30
  [arXiv:1012.1004 [nucl-ex]];
  %%CITATION = PHLTA,B696,30;%%
% 
%\cite{Aamodt:2010cz}
% \bibitem{Aamodt:2010cz}
%   K.~Aamodt {\it et al.}  [ALICE Collaboration],
  %``Centrality dependence of the charged-particle multiplicity density at
  %mid-rapidity in Pb-Pb collisions at sqrt(sNN) = 2.76 TeV,''
  Phys.\ Rev.\ Lett.\  {\bf 106}, 032301 (2011)
  [arXiv:1012.1657 [nucl-ex]];
  %%CITATION = PRLTA,106,032301;%%
% 
%\cite{Aamodt:2011mr}
% \bibitem{Aamodt:2011mr}
%   K.~Aamodt {\it et al.}  [ALICE Collaboration],
  %``Two-pion Bose-Einstein correlations in central PbPb collisions at
  %sqrt(s_NN) = 2.76 TeV,''
  Phys.\ Lett.\  B {\bf 696}, 328 (2011)
  [arXiv:1012.4035 [nucl-ex]].
  %%CITATION = PHLTA,B696,328;%%

%\cite{first_deconfinement_prediction}
\bibitem{first_deconfinement_prediction}
  N.~Cabibbo and G.~Parisi,
  %``Exponential Hadronic Spectrum And Quark Liberation,''
  Phys.\ Lett.\  B {\bf 59} (1975) 67.
  %%CITATION = PHLTA,B59,67;%%
  J.~C.~Collins and M.~J.~Perry,
  %``Superdense Matter: Neutrons Or Asymptotically Free Quarks?,''
  Phys.\ Rev.\ Lett.\  {\bf 34}, 1353 (1975).
  %%CITATION = PRLTA,34,1353;%%

%\cite{perturbative_computations}
\bibitem{perturbative_computations}
  P.~B.~Arnold and C.-X.~Zhai,
  %``The Three loop free energy for pure gauge QCD,''
  Phys.\ Rev.\  {\bf D50 } (1994)  7603
  [hep-ph/9408276];
  %``The Three loop free energy for high temperature QED and QCD with fermions,''
  Phys.\ Rev.\  {\bf D51 } (1995)  1906
  [hep-ph/9410360].
  C.-X.~Zhai and B.~M.~Kastening,
  %``The Free energy of hot gauge theories with fermions through g**5,''
  Phys.\ Rev.\  {\bf D52 } (1995)  7232
  [hep-ph/9507380].
  J.~O.~Andersen, E.~Braaten and M.~Strickland,
  %``Hard thermal loop resummation of the free energy of a hot gluon plasma,''
  Phys.\ Rev.\ Lett.\  {\bf 83 } (1999)  2139
  [hep-ph/9902327].
  J.-P.~Blaizot, E.~Iancu and A.~Rebhan,
  %``Approximately selfconsistent resummations for the thermodynamics of the quark gluon plasma. 1. Entropy and density,''
  Phys.\ Rev.\  {\bf D63 } (2001)  065003
  [hep-ph/0005003].
  K.~Kajantie, M.~Laine, K.~Rummukainen and Y.~Schr\"oder,
  %``The Pressure of hot QCD up to g6 ln(1/g),''
  Phys.\ Rev.\  {\bf D67 } (2003)  105008
  [hep-ph/0211321].
  M.~Laine and Y.~Schr\"oder,
  %``Two-loop QCD gauge coupling at high temperatures,''
  JHEP {\bf 0503 } (2005)  067
  [hep-ph/0503061].
  A.~Hietanen, K.~Kajantie, M.~Laine, K.~Rummukainen and Y.~Schr\"oder,
  %``Three-dimensional physics and the pressure of hot QCD,''
  Phys.\ Rev.\  {\bf D79 } (2009)  045018
  [arXiv:0811.4664 [hep-lat]].
  J.~O.~Andersen, M.~Strickland and N.~Su,
  %``Gluon Thermodynamics at Intermediate Coupling,''
  Phys.\ Rev.\ Lett.\  {\bf 104 } (2010)  122003
  [arXiv:0911.0676 [hep-ph]].
  J.~O.~Andersen, L.~E.~Leganger, M.~Strickland and N.~Su,
  %``NNLO hard-thermal-loop thermodynamics for QCD,''
  Phys.\ Lett.\  {\bf B696 } (2011)  468
  [arXiv:1009.4644 [hep-ph]].

%\cite{Linde_problem}
\bibitem{Linde_problem}
  A.~D.~Linde,
  %``Infrared Problem In Thermodynamics Of The Yang-Mills Gas,''
  Phys.\ Lett.\  B {\bf 96}, 289 (1980).
  %%CITATION = PHLTA,B96,289;%%
  D.~J.~Gross, R.~D.~Pisarski and L.~G.~Yaffe,
  %``QCD And Instantons At Finite Temperature,''
  Rev.\ Mod.\ Phys.\  {\bf 53}, 43 (1981).
  %%CITATION = RMPHA,53,43;%%

%\cite{finite_T_lattice_reviews}
\bibitem{finite_T_lattice_reviews}
  C.~DeTar and U.~M.~Heller,
  %``QCD Thermodynamics from the Lattice,''
  Eur.\ Phys.\ J.\  {\bf A41 } (2009)  405
  [arXiv:0905.2949 [hep-lat]].
  M.~Laine,
  %``Finite-temperature QCD,''
  PoS {\bf LAT2009 } (2009)  006
  [arXiv:0910.5168 [hep-lat]].
  K.~Kanaya,
  %``Finite Temperature QCD on the Lattice -- Status 2010,''
  PoS {\bf LATTICE2010 } (2010)  012
  [arXiv:1012.4247 [hep-lat]].
  C.~DeTar,
  %``QCD Thermodynamics on the Lattice: Recent Results,''
  [arXiv:1101.0208 [hep-lat]].
  %%CITATION = ARXIV:1101.0208;%%
  L.~Levkova,
  %``QCD at finite temperature and density,''
  PoS {\bf LATTICE2011 } (2011)  012.

%\cite{finite_density_lattice_QCD}
\bibitem{finite_density_lattice_QCD}
  S.~Gupta,
  %``QCD at finite density,''
  PoS {\bf LATTICE2010 } (2010)  007
  [arXiv:1101.0109 [hep-lat]].

%\cite{DePietri:2007ak}
\bibitem{DePietri:2007ak}
  R.~De Pietri, A.~Feo, E.~Seiler and I.~O.~Stamatescu,
  %``A Model for QCD at high density and large quark mass,''
  Phys.\ Rev.\  D {\bf 76} (2007) 114501
  [arXiv:0705.3420 [hep-lat]].
  %%CITATION = PHRVA,D76,114501;%%

%\cite{Yaffe:1981vf}
\bibitem{Yaffe:1981vf}
  L.~G.~Yaffe,
  %``Large N Limits As Classical Mechanics,''
  Rev.\ Mod.\ Phys.\  {\bf 54} (1982) 407.
  %%CITATION = RMPHA,54,407;%%

\bibitem{'tHooftlargeN}
  G.~'t Hooft,
  %``A Planar Diagram Theory for Strong Interactions,''
  Nucl.\ Phys.\ {\bf B72} (1974) 461.
  %%CITATION = NUPHA,B72,461;%%
  E.~Witten,
  %``Baryons in the 1/N Expansion,''
  Nucl.\ Phys.\ {\bf B160 } (1979) 57.
  %%CITATION = NUPHA,B160,57;%%
  A.~V.~Manohar,
  %``Large N QCD,''  
  arXiv:hep-ph/9802419.
  %%CITATION = HEP-PH/9802419;%%
  Y.~Makeenko,
  %``Large-N gauge theories,''
  arXiv:hep-th/0001047.
  %%CITATION = HEP-TH/0001047;%%

%\cite{OZI_rule}
\bibitem{OZI_rule}
  S.~Okubo,
  %``Phi meson and unitary symmetry model,''
  Phys.\ Lett.\  {\bf 5 } (1963)  165.
  G.~Zweig,
  %``An SU(3) model for strong interaction symmetry and its breaking,''
  in D.~B.~Lichtenberg and S.~P.~Rosen (editors) \emph{Developments in the quark theory of hadrons}, {\bf 1} 22, and CERN Geneva-TH. 401 (1964), 24 p.
  J.~Iizuka,
  %``Systematics and phenomenology of meson family,''
  Prog.\ Theor.\ Phys.\ Suppl.\  {\bf 37 } (1966)  21.

\bibitem{Aharony:1999ti}
  O.~Aharony, S.~S.~Gubser, J.~M.~Maldacena, H.~Ooguri and Y.~Oz,
  %``Large N field theories, string theory and gravity,''
  Phys.\ Rept.\ {\bf 323 } (2000) 183
  [hep-th/9905111].
  %%CITATION = HEP-TH/9905111;%%

%\cite{Gopakumar:1994iq}
\bibitem{Gopakumar:1994iq}
  R.~Gopakumar and D.~J.~Gross,
  %``Mastering the master field,''
  Nucl.\ Phys.\  {\bf B451 } (1995)  379
  [hep-th/9411021].

\bibitem{volume_reduction}
  T.~Eguchi and H.~Kawai,
  %``Reduction of Dynamical Degrees of Freedom in the Large N Gauge Theory,''
  Phys.\ Rev.\ Lett.\ {\bf 48 } (1982)  1063.
  %%CITATION = PRLTA,48,1063;%%
  A.~Gonz\'alez-Arroyo and M.~Okawa,
  %``A Twisted Model For Large N Lattice Gauge Theory,''
  Phys.\ Lett.\ {\bf B120 } (1983) 174.
  %%CITATION = PHLTA,B120,174;%%
  G.~Bhanot, U.~M.~Heller and H.~Neuberger,
  %``The Quenched Eguchi-Kawai Model,''
  Phys.\ Lett.\ {\bf B113 } (1982)  47.
  %%CITATION = PHLTA,B113,47;%%
  P.~Kovtun, M.~\"Unsal and L.~G.~Yaffe,
  %``Volume independence in large N(c) QCD-like gauge theories,''
  JHEP {\bf 0706 } (2007)  019
  [hep-th/0702021].
  %%CITATION = HEP-TH/0702021;%%
  M.~\"Unsal,
  %``Large-N volume independence, conformality and confinement,''
  PoS {\bf LATTICE2011} (2011)  003.
  %%CITATION = POSCI,LATTICE2011,003;%%

\bibitem{McLerran:2007qj}
  L.~McLerran and R.~D.~Pisarski,
  %``Phases of cold, dense quarks at large N(c),''
  Nucl.\ Phys.\ {\bf A796 } (2007)  83
  [arXiv:0706.2191 [hep-ph]].
  %%CITATION = ARXIV:0706.2191;%%

\bibitem{Maldacena_conjecture}
%\cite{Maldacena:1997re}
% \bibitem{Maldacena:1997re}
  J.~M.~Maldacena,
  %``The large N limit of superconformal field theories and supergravity,''
  Adv.\ Theor.\ Math.\ Phys.\  {\bf 2}, 231 (1998)
  [Int.\ J.\ Theor.\ Phys.\  {\bf 38}, 1113 (1999)]
  [arXiv:hep-th/9711200].
  %%CITATION = IJTPB,38,1113;%%
% 
%\cite{Gubser:1998bc}
% \bibitem{Gubser:1998bc}
  S.~S.~Gubser, I.~R.~Klebanov and A.~M.~Polyakov,
  %``Gauge theory correlators from non-critical string theory,''
  Phys.\ Lett.\  B {\bf 428}, 105 (1998)
  [arXiv:hep-th/9802109].
  %%CITATION = PHLTA,B428,105;%%
% 
%\cite{Witten:1998qj}
% \bibitem{Witten:1998qj}
  E.~Witten,
  %``Anti-de Sitter space and holography,''
  Adv.\ Theor.\ Math.\ Phys.\  {\bf 2}, 253 (1998)
  [arXiv:hep-th/9802150].
  %%CITATION = 00203,2,253;%%

%\cite{gauge_string_applications_to_QCD_plasma}
\bibitem{gauge_string_applications_to_QCD_plasma}
  D.~T.~Son and A.~O.~Starinets,
  %``Viscosity, Black Holes, and Quantum Field Theory,''
  Ann.\ Rev.\ Nucl.\ Part.\ Sci.\  {\bf 57}, 95 (2007)
  [arXiv:0704.0240 [hep-th]].
  %%CITATION = ARNUA,57,95;%%
  D.~Mateos,
  %``String Theory and Quantum Chromodynamics,''
  Class.\ Quant.\ Grav.\  {\bf 24}, S713 (2007)
  [arXiv:0709.1523 [hep-th]].
  %%CITATION = CQGRD,24,S713;%%
  S.~S.~Gubser and A.~Karch,
  %``From gauge-string duality to strong interactions: a Pedestrian's Guide,''
  Ann.\ Rev.\ Nucl.\ Part.\ Sci.\  {\bf 59}, 145 (2009)
  [arXiv:0901.0935 [hep-th]].
  %%CITATION = ARNUA,59,145;%%
  J.~Casalderrey-Solana, H.~Liu, D.~Mateos, K.~Rajagopal and U.~A.~Wiedemann,
  %``Gauge/String Duality, Hot QCD and Heavy Ion Collisions,''  
  arXiv:1101.0618 [hep-th].

\bibitem{finiteTlargeNlatticeresults}
%\cite{Lucini:2002ku}
% \bibitem{Lucini:2002ku}
  B.~Lucini, M.~Teper and U.~Wenger,
  %``The deconfinement transition in SU(N) gauge theories,''
  Phys.\ Lett.\  B {\bf 545}, 197 (2002)
  [arXiv:hep-lat/0206029];
  %%CITATION = PHLTA,B545,197;%%
% 
%\cite{Lucini:2003zr}
% \bibitem{Lucini:2003zr}
%   B.~Lucini, M.~Teper and U.~Wenger,
  %``The high temperature phase transition in SU(N) gauge theories,''
  JHEP {\bf 0401}, 061 (2004)
  [arXiv:hep-lat/0307017];
  %%CITATION = JHEPA,0401,061;%%
% 
%\cite{Lucini:2004yh}
% \bibitem{Lucini:2004yh}
%   B.~Lucini, M.~Teper and U.~Wenger,
  %``Topology of SU(N) gauge theories at T approx. 0 and T approx. T(c),''
  Nucl.\ Phys.\  B {\bf 715}, 461 (2005)
  [arXiv:hep-lat/0401028];
  %%CITATION = NUPHA,B715,461;%%
% 
%\cite{Lucini:2005vg}
% \bibitem{Lucini:2005vg}
%   B.~Lucini, M.~Teper and U.~Wenger,
  %``Properties of the deconfining phase transition in SU(N) gauge theories,''
  JHEP {\bf 0502}, 033 (2005)
  [arXiv:hep-lat/0502003].
  %%CITATION = JHEPA,0502,033;%%
% 
%\cite{Bringoltz:2005rr}
% \bibitem{Bringoltz:2005rr}
  B.~Bringoltz and M.~Teper,
  %``The pressure of the SU(N) lattice gauge theory at large-N,''
  Phys.\ Lett.\  B {\bf 628} (2005) 113
  [arXiv:hep-lat/0506034];
  %%CITATION = PHLTA,B628,113;%%
% 
%\cite{Bringoltz:2005xx}
% \bibitem{Bringoltz:2005xx}
%   B.~Bringoltz and M.~Teper,
  %``In search of a Hagedorn transition in SU(N) lattice gauge theories at
  %large-N,''
  Phys.\ Rev.\  D {\bf 73} (2006) 014517
  [arXiv:hep-lat/0508021];
  %%CITATION = PHRVA,D73,014517;%%
% 
%\cite{Datta:2009jn}
% \bibitem{Datta:2009jn}
  S.~Datta and S.~Gupta,
  %``Scaling and the continuum limit of the finite temperature deconfinement
  %transition in SU($N_c$) pure gauge theory,''
  Phys.\ Rev.\  D {\bf 80} (2009) 114504
  [arXiv:0909.5591 [hep-lat]];
%\cite{Datta:2010sq}
% \bibitem{Datta:2010sq}
%   S.~Datta and S.~Gupta,
  %``Continuum Thermodynamics of the GluoN_c Plasma,''
  Phys.\ Rev.\  D {\bf 82} (2010) 114505
  [arXiv:1006.0938 [hep-lat]].
  %%CITATION = PHRVA,D82,114505;%%
% 
%\cite{Mykkanen:2012ri}
% \bibitem{Mykkanen:2012ri}
  A.~Mykk\"anen, M.~Panero and K.~Rummukainen,
  %``Casimir scaling and renormalization of Polyakov loops in large-N gauge theories,''
  JHEP {\bf 1205} (2012) 069
  [arXiv:1202.2762 [hep-lat]];
  %%CITATION = ARXIV:1202.2762;%%
%\cite{Mykkanen:2011kz}
% \bibitem{Mykkanen:2011kz}
%   A.~Mykk\"anen, M.~Panero and K.~Rummukainen,
  %``Renormalization of Polyakov loops in different representations and the
  %large-N limit,''
  PoS LATTICE {\bf 2011} (2011) 211
  [arXiv:1110.3146 [hep-lat]].
  %%CITATION = ARXIV:1110.3146;%%
% 
%\cite{Lucini:2012wq}
% \bibitem{Lucini:2012wq}
  B.~Lucini, A.~Rago and E.~Rinaldi,
  %``SU(N_c) gauge theories at deconfinement,''
  Phys.\ Lett.\ B {\bf 712} (2012) 279
  [arXiv:1202.6684 [hep-lat]].
  %%CITATION = ARXIV:1202.6684;%%

%\cite{Panero:2009tv}
\bibitem{Panero:2009tv}
  M.~Panero,
  %``Thermodynamics of the QCD plasma and the large-N limit,''
  Phys.\ Rev.\ Lett.\  {\bf 103}, 232001 (2009)
  [arXiv:0907.3719 [hep-lat]];
  %%CITATION = PRLTA,103,232001;%%
% 
%\cite{Panero:2009wr}
% \bibitem{Panero:2009wr}
%   M.~Panero,
  %``Thermodynamics of the strongly interacting gluon plasma in the large-N
  %limit,''
  PoS {\bf LAT2009}, 172 (2009)
  [arXiv:0912.2448 [hep-lat]].
  %%CITATION = POSCI,LAT2009,172;%%

\bibitem{SU3_EoS}
  G.~Boyd, J.~Engels, F.~Karsch, E.~Laermann, C.~Legeland, M.~L\"utgemeier and B.~Petersson,
  %``Thermodynamics of SU(3) lattice gauge theory,''
  Nucl.\ Phys.\  B {\bf 469}, 419 (1996)
  [arXiv:hep-lat/9602007].
  %%CITATION = NUPHA,B469,419;%% 
  S.~Bors\'anyi, G.~Endr\H{o}di, Z.~Fodor, S.~D.~Katz and K.~K.~Szab\'o,
  %``Lattice SU(3) thermodynamics and the onset of perturbative behaviour,''
  arXiv:1104.0013 [hep-ph].
  %%CITATION = ARXIV:1104.0013;%%

%\cite{Teper:2009uf}
\bibitem{Teper:2009uf}
  M.~Teper,
  %``Large N and confining flux tubes as strings - a view from the lattice,''
  Acta Phys.\ Polon.\  B {\bf 40} (2009) 3249
  [arXiv:0912.3339 [hep-lat]].
  %%CITATION = APPOA,B40,3249;%%

%\cite{Gross:1980he}
\bibitem{Gross:1980he}
  D.~J.~Gross and E.~Witten,
  %``Possible Third Order Phase Transition in the Large N Lattice Gauge Theory,''
  Phys.\ Rev.\  {\bf D21 } (1980)  446.

%\cite{Teper:1998te}
\bibitem{Teper:1998te}
  M.~J.~Teper,
  %``SU(N) gauge theories in 2+1 dimensions,''
  Phys.\ Rev.\  D {\bf 59}, 014512 (1999)
  [arXiv:hep-lat/9804008].
  %%CITATION = PHRVA,D59,014512;%%
  R.~W.~Johnson and M.~J.~Teper,
  %``String models of glueballs and the spectrum of SU(N) gauge theories in
  %(2+1)-dimensions,''
  Phys.\ Rev.\  D {\bf 66} (2002) 036006
  [arXiv:hep-ph/0012287].
  %%CITATION = PHRVA,D66,036006;%%
  H.~B.~Meyer and M.~J.~Teper,
  %``Glueball Regge trajectories in (2+1)-dimensional gauge theories,''
  Nucl.\ Phys.\  {\bf B668 }, 111 (2003)
  [hep-lat/0306019].
  %%CITATION = HEP-LAT/0306019;%%

%\cite{Sachdev:2010ch}
\bibitem{Sachdev:2010ch}
  S.~Sachdev,
  %``Condensed matter and AdS/CFT,''  
  arXiv:1002.2947 [hep-th].

%\cite{Tsquare_in_4D}
\bibitem{Tsquare_in_4D}
  P.~N.~Meisinger, T.~R.~Miller and M.~C.~Ogilvie,
  %``Phenomenological equations of state for the quark gluon plasma,''
  Phys.\ Rev.\  {\bf D65 } (2002)  034009
  [hep-ph/0108009].
  A.~Peshier, B.~K\"ampfer, O.~P.~Pavlenko and G.~Soff,
  %``A Massive Quasiparticle Model Of The SU(3) Gluon Plasma,''
  Phys.\ Rev.\  D {\bf 54}, 2399 (1996).
  %%CITATION = PHRVA,D54,2399;%%
  E.~Meg\'{\i}as, E.~Ruiz Arriola and L.~L.~Salcedo,
  %``Dimension two condensates and the Polyakov loop above the deconfinement phase transition,''
  JHEP {\bf 0601 } (2006)  073
  [hep-ph/0505215];
  %``Trace Anomaly, Thermal Power Corrections and Dimension Two condensates in the deconfined phase,''
  Phys.\ Rev.\  {\bf D80 } (2009)  056005
  [arXiv:0903.1060 [hep-ph]].
  O.~Andreev,
  %``1/q**2 corrections and gauge/string duality,''
  Phys.\ Rev.\  {\bf D73 } (2006)  107901
  [hep-th/0603170];
  %``Some Thermodynamic Aspects of Pure Glue, Fuzzy Bags and Gauge/String Duality,''
  Phys.\ Rev.\  {\bf D76 } (2007)  087702
  [arXiv:0706.3120 [hep-ph]].
  R.~D.~Pisarski,
  %``Fuzzy Bags and Wilson Lines,''
  Prog.\ Theor.\ Phys.\ Suppl.\  {\bf 168 } (2007)  276
  [hep-ph/0612191].
  F.~Brau and F.~Buisseret,
  %``Glueballs and statistical mechanics of the gluon plasma,''
  Phys.\ Rev.\  {\bf D79 } (2009)  114007
  [arXiv:0902.4836 [hep-ph]].
  F.~Buisseret and G.~Lacroix,
  %``A Minimal quasiparticle approach for the QGP and its large-$N_c$ limits,''
  Eur.\ Phys.\ J.\  {\bf C70 } (2010)  1051
  [arXiv:1006.0655 [hep-ph]].

%\cite{Caselle:2011fy}
\bibitem{Caselle:2011fy}
  M.~Caselle, L.~Castagnini, A.~Feo, F.~Gliozzi and M.~Panero,
  %``Thermodynamics of SU(N) Yang-Mills theories in 2+1 dimensions I - The
  %confining phase,''
  JHEP {\bf 1106} (2011) 142
  [arXiv:1105.0359 [hep-lat]];
  %``Thermodynamics of SU(N) gauge theories in 2+1 dimensions in the $T <\ T_c$ regime,''
  PoS {\bf LATTICE2010 } (2010)  184
  [arXiv:1011.4883 [hep-lat]].
  %%CITATION = POSCI,LATTICE2010,184;%%

 %\cite{Witten:1998zw}
\bibitem{Witten2}
  E.~Witten,
  %``Anti-de Sitter space, thermal phase transition, and confinement in  gauge theories,''
  Adv.\ Theor.\ Math.\ Phys.\  {\bf 2} (1998) 505.
   %%CITATION = 00203,2,505;%%

\bibitem{pheno}
  J.~Erlich, E.~Katz, D.~T.~Son and M.~A.~Stephanov,
  %``QCD and a Holographic Model of Hadrons,''
  Phys.\ Rev.\ Lett.\  {\bf 95} (2005) 261602
  [hep-ph/0501128].
  L.~Da Rold and A.~Pomarol,
  %``Chiral symmetry breaking from five dimensional spaces,''
  Nucl.\ Phys.\  B {\bf 721} (2005) 79
  [hep-ph/0501218].

\bibitem{GK}
  U.~G\"ursoy and E.~Kiritsis,
  %``Exploring improved holographic theories for QCD: Part I,''
  JHEP {\bf 0802} (2008) 032
  [arXiv:0707.1324 [hep-th]].

\bibitem{GKN}
  U.~G\"ursoy, E.~Kiritsis and F.~Nitti,
  %``Exploring improved holographic theories for QCD: Part II,''
  JHEP {\bf 0802} (2008) 019
  [arXiv:0707.1349 [hep-th]].

\bibitem{Kiritsis}
  U.~G\"ursoy, E.~Kiritsis, L.~Mazzanti, G.~Michalogiorgakis, F.~Nitti,
  %``Improved Holographic QCD,''
  Lect.\ Notes Phys.\  {\bf 828 } (2011)  79
  [arXiv:1006.5461 [hep-th]].

%\cite{Gubser:2008ny}
\bibitem{Gubser}
  S.~S.~Gubser and A.~Nellore,
  %``Mimicking the QCD equation of state with a dual black hole,''
  Phys.\ Rev.\  {\bf D78 } (2008)  086007
  [arXiv:0804.0434 [hep-th]].

\bibitem{GKMN2}
  U.~G\"ursoy, E.~Kiritsis, L.~Mazzanti and F.~Nitti,
  %``Holography and Thermodynamics of 5D Dilaton-gravity,''
  JHEP {\bf 0905} (2009) 033
  [arXiv:0812.0792 [hep-th]].
  %%CITATION = JHEPA,0905,033;%%

\bibitem{G1}
  U.~G\"ursoy,
  %``Gravity/Spin-model correspondence and holographic superfluids,''
  JHEP {\bf 1012} (2010) 062
  [arXiv:1007.4854 [hep-th]].
  %%CITATION = JHEPA,1012,062;%%

\bibitem{GKMN1}
  U.~G\"ursoy, E.~Kiritsis, L.~Mazzanti and F.~Nitti,
  %``Deconfinement and Gluon Plasma Dynamics in Improved Holographic QCD,''
  Phys.\ Rev.\ Lett.\  {\bf 101} (2008) 181601
  [arXiv:0804.0899 [hep-th]].

\bibitem{GKMN3}
  U.~G\"ursoy, E.~Kiritsis, L.~Mazzanti and F.~Nitti,
  %``Improved Holographic Yang-Mills at Finite Temperature: Comparison with
  %Data,''
  Nucl.\ Phys.\  B {\bf 820} (2009) 148
  [arXiv:0903.2859 [hep-th]].
  %%CITATION = NUPHA,B820,148;%%

%\cite{Kajantie}
\bibitem{Kajantie}
  J.~Alanen, K.~Kajantie and V.~Suur-Uski,
  %``A gauge/gravity duality model for gauge theory thermodynamics,''
  Phys.\ Rev.\  D {\bf 80}, 126008 (2009)
  [arXiv:0911.2114 [hep-ph]];
  %%CITATION = PHRVA,D80,126008;%%
  %``Spatial string tension of finite temperature QCD matter in gauge/gravity duality,''
  Phys.\ Rev.\  {\bf D80 } (2009)  075017
  [arXiv:0905.2032 [hep-ph]].
  E.~Meg\'{\i}as, H.~J.~Pirner and K.~Veschgini,
  %``QCD thermodynamics using five-dimensional gravity,''
  Phys.\ Rev.\  {\bf D83 } (2011)  056003
  [arXiv:1009.2953 [hep-ph]].
  K.~Kajantie, M.~Kr\v s\v s\'ak, M.~Veps\"al\"ainen and A.~Vuorinen,
  %``Frequency and wave number dependence of the shear correlator in strongly coupled hot Yang-Mills theory,''  
  arXiv:1104.5352 [hep-ph].
  J.~Alanen, T.~Alho, K.~Kajantie and K.~Tuominen,
  %``Mass spectrum and thermodynamics of quasi-conformal gauge theories from gauge/gravity duality,''  
  arXiv:1107.3362 [hep-th].

\bibitem{G2}
  U.~G\"ursoy,
  %``Continuous Hawking-Page transitions in Einstein-scalar gravity,''
  JHEP {\bf 1101} (2011) 086
  [arXiv:1007.0500 [hep-th]].
  %%CITATION = JHEPA,1101,086;%%

%\cite{Liddle:2008kk}
\bibitem{Liddle:2008kk}
  J.~Liddle and M.~Teper,
  %``The deconfining phase transition in D=2+1 SU(N) gauge theories,''
  arXiv:0803.2128 [hep-lat].
  %%CITATION = ARXIV:0803.2128;%%

%\cite{Bialas:2008rk}
\bibitem{Bialas:2008rk}
  P.~Bialas, L.~Daniel, A.~Morel and B.~Petersson,
  %``Thermodynamics of SU(3) Gauge Theory in 2 + 1 Dimensions,''
  Nucl.\ Phys.\  {\bf B807 } (2009)  547
  [arXiv:0807.0855 [hep-lat]].
  %%CITATION = ARXIV:0807.0855;%%

\bibitem{SUN_thermodynamics_in_2_plus_1_dimensions}
  P.~de Forcrand and O.~Jahn,
  %``Deconfinement transition in 2+1-dimensional SU(4) lattice gauge theory,''
  Nucl.\ Phys.\ Proc.\ Suppl.\  {\bf 129} (2004) 709
  [arXiv:hep-lat/0309153].
  %%CITATION = NUPHZ,129,709;%%
  K.~Holland, M.~Pepe and U.-J.~Wiese,
  % ``Revisiting the deconfinement phase transition in SU(4) Yang-Mills theory in
  %2+1 dimensions,''
  JHEP {\bf 0802}, 041 (2008)
  [arXiv:0712.1216 [hep-lat]].
  %%CITATION = JHEPA,0802,041;%%
  K.~Holland,
  %``Another weak first order deconfinement transition: Three-dimensional  SU(5)
  %gauge theory,''
  JHEP {\bf 0601}, 023 (2006)
  [arXiv:hep-lat/0509041].
  %%CITATION = JHEPA,0601,023;%%
  L.~von Smekal, S.~R.~Edwards and N.~Strodthoff,
  %``Universal Aspects of Deconfinement in 2+1 Dimensions,''
  AIP Conf.\ Proc.\  {\bf 1343 } (2011)  212
  [arXiv:1012.1712 [hep-ph]].
  %%CITATION = ARXIV:1012.1712;%%

%\cite{Gubser:1998nz}
\bibitem{Gubser:1998nz}
  S.~S.~Gubser, I.~R.~Klebanov and A.~A.~Tseytlin,
  %``Coupling constant dependence in the thermodynamics of N=4 supersymmetric Yang-Mills theory,''
  Nucl.\ Phys.\  {\bf B534 } (1998)  202
  [hep-th/9805156].

\bibitem{3d_YM_renormalization_properties}
%\cite{Reuter:1993nn}
% \bibitem{Reuter:1993nn}
  M.~Reuter and C.~Wetterich,
  %``Running gauge coupling in three-dimensions and the electroweak phase
  %transition,''
  Nucl.\ Phys.\  B {\bf 408}, 91 (1993).
  %%CITATION = NUPHA,B408,91;%%
% 
%\cite{Farakos:1994xh}
% \bibitem{Farakos:1994xh}
  K.~Farakos, K.~Kajantie, K.~Rummukainen and M.~E.~Shaposhnikov,
  %``3-d physics and the electroweak phase transition: A Framework for lattice
  %Monte Carlo analysis,''
  Nucl.\ Phys.\  B {\bf 442} (1995) 317
  [arXiv:hep-lat/9412091].
  %%CITATION = NUPHA,B442,317;%%

%\cite{Wilson:1974sk}
\bibitem{Wilson:1974sk}
  K.~G.~Wilson,
  %``Confinement of Quarks,''
  Phys.\ Rev.\  {\bf D10 } (1974)  2445.

%\cite{Caselle:2004er}
\bibitem{Caselle:2004er}
  M.~Caselle, M.~Pepe and A.~Rago,
  %``Static quark potential and effective string corrections in the (2+1)-d
  %SU(2) Yang-Mills theory,''
  JHEP {\bf 0410} (2004) 005
  [arXiv:hep-lat/0406008].
  %%CITATION = JHEPA,0410,005;%%

%\cite{lattice_perturbation_theory}
\bibitem{lattice_perturbation_theory}
  B.~All\'es, M.~Campostrini, A.~Feo and H.~Panagopoulos,
  %``The Three loop lattice free energy,''
  Phys.\ Lett.\  {\bf B324 } (1994)  433
  [hep-lat/9306001];
  %%CITATION = ARXIV:HEP-LAT/9306001;%%
  %``Lattice perturbation theory by computer algebra: A Three loop result for the topological susceptibility,''
  Nucl.\ Phys.\  {\bf B413 } (1994)  553
  [hep-lat/9301012].
  %%CITATION = ARXIV:HEP-LAT/9301012;%%
  C.~Christou, A.~Feo, H.~Panagopoulos and E.~Vicari,
  %``The Three loop Beta function of SU(N) lattice gauge theories with Wilson fermions,''
  Nucl.\ Phys.\  {\bf B525 } (1998)  387
  [hep-lat/9801007].
  %%CITATION = ARXIV:HEP-LAT/9801007;%%
  B.~All\'es, A.~Feo and H.~Panagopoulos,
  %``The Three loop Beta function in SU(N) lattice gauge theories,''
  Nucl.\ Phys.\  {\bf B491 } (1997)  498
  [hep-lat/9609025];
  %%CITATION = ARXIV:HEP-LAT/9609025;%%
  %``Asymptotic scaling corrections in QCD with Wilson fermions from the three loop average plaquette,''
  Phys.\ Lett.\  {\bf B426 } (1998)  361
  [hep-lat/9801003].
  %%CITATION = ARXIV:HEP-LAT/9801003;%%

\bibitem{algorithm}
  M.~Creutz,
  %``Monte Carlo Study Of Quantized SU(2) Gauge Theory,''
  Phys.\ Rev.\  D {\bf 21}, 2308 (1980).
  %%CITATION = PHRVA,D21,2308;%%
  A.~D.~Kennedy and B.~J.~Pendleton,
  %``Improved Heat Bath Method For Monte Carlo Calculations In Lattice Gauge
  %Theories,''
  Phys.\ Lett.\  B {\bf 156}, 393 (1985).
  %%CITATION = PHLTA,B156,393;%%
  S.~L.~Adler,
  %``AN OVERRELAXATION METHOD FOR THE MONTE CARLO EVALUATION OF THE PARTITION
  %FUNCTION FOR MULTIQUADRATIC ACTIONS,''
  Phys.\ Rev.\  D {\bf 23} (1981) 2901.
  %%CITATION = PHRVA,D23,2901;%%
  F.~R.~Brown and T.~J.~Woch,
  %``Overrelaxed Heat Bath and Metropolis Algorithms for Accelerating Pure Gauge
  %Monte Carlo Calculations,''
  Phys.\ Rev.\ Lett.\  {\bf 58} (1987) 2394.
  %%CITATION = PRLTA,58,2394;%%
  N.~Cabibbo and E.~Marinari,
  %``A New Method For Updating SU(N) Matrices In Computer Simulations Of Gauge
  %Theories,''
  Phys.\ Lett.\  B {\bf 119}, 387 (1982).
  %%CITATION = PHLTA,B119,387;%%

%\cite{Edwards:2004sx}
\bibitem{Edwards:2004sx}
  R.~G.~Edwards and B.~Jo\'o [SciDAC Collaboration and LHPC Collaboration and UKQCD Collaboration],
  %``The Chroma software system for lattice QCD,''
  Nucl.\ Phys.\ Proc.\ Suppl.\  {\bf 140} (2005) 832
  [arXiv:hep-lat/0409003].
  %%CITATION = NUPHZ,140,832;%%

%\cite{Gliozzi_finite_volume}
\bibitem{Gliozzi_finite_volume}
  F.~Gliozzi,
  %``The Stefan-Boltzmann law in a small box and the pressure deficit in hot
  %SU(N) lattice gauge theory,''
  J.\ Phys.\ A  {\bf 40}, F375 (2007)
  [arXiv:hep-lat/0701020].
  %%CITATION = JPAGB,A40,F375;%%
  M.~Panero,
  %``Geometric effects in lattice QCD thermodynamics,''
  PoS {\bf LATTICE2008}, 175 (2008)
  [arXiv:0808.1672 [hep-lat]].
  %%CITATION = POSCI,LATTICE2008,175;%%

%\cite{Engels:1990vr}
\bibitem{Engels:1990vr}
  J.~Engels, J.~Fingberg, F.~Karsch, D.~Miller and M.~Weber,
  %``Nonperturbative thermodynamics of SU(N) gauge theories,''
  Phys.\ Lett.\  B {\bf 252} (1990) 625.
  %%CITATION = PHLTA,B252,625;%%

%\cite{Caselle:2007yc}
\bibitem{Caselle:2007yc}
  M.~Caselle, M.~Hasenbusch and M.~Panero,
  %``The interface free energy: Comparison of accurate Monte Carlo results for
  %the 3D Ising model with effective interface models,''
  JHEP {\bf 0709} (2007) 117
  [arXiv:0707.0055 [hep-lat]].
  %%CITATION = JHEPA,0709,117;%%

%\cite{Engels:1999tk}
\bibitem{Engels:1999tk}
  J.~Engels, F.~Karsch and T.~Scheideler,
  %``Determination of anisotropy coefficients for SU(3) gauge actions from the integral and matching methods,''
  Nucl.\ Phys.\  {\bf B564 } (2000)  303
  [hep-lat/9905002].
  T.~Scheideler,
  Ph.D. thesis, University of Bielefeld (1998), available for download from:\\
  \verb+http://www2.physik.uni-bielefeld.de/3113.html+
%   \verb+http://www2.physik.uni-bielefeld.de/fileadmin/user_upload/theory_e6/PhD_Theses/scheideler.ps.gz+

%\cite{Elze:1988zs}
\bibitem{Elze:1988zs}
  H.~T.~Elze, K.~Kajantie and J.~I.~Kapusta,
  %``Screening and plasmon in QCD on a finite lattice,''
  Nucl.\ Phys.\  B {\bf 304} (1988) 832.
  %%CITATION = NUPHA,B304,832;%%

\end{thebibliography}
\end{document}